# All-Silicon Topological Semimetals with Closed Nodal Line

Zhifeng Liu,[1,2] Hongli Xin,[1] Li Fu,[1] Yingqiao Liu,[3] Tielei Song,[1] Xin Cui,[1] Guojun Zhao,[1] Jijun Zhao[2,3*]

[1]*School of Physical Science and Technology, Inner Mongolia University, Hohhot 010021, China*

[2]*Beijing Computational Science Research Center, Beijing 100094, China*

[3]*Key Laboratory of Materials Modification by Laser, Ion and Electron Beams* (*Dalian University of Technology*), *Ministry of Education, Dalian 116024, China*

Owing to the natural compatibility with current semiconductor industry, silicon allotropes with diverse structural and electronic properties provide promising platforms for the next-generation Si-based devices. After screening 230 all-silicon crystals in the zeolite frameworks by first-principles calculations, we disclose two structurally stable Si allotropes (AHT-$Si_{24}$ and VFI-$Si_{36}$) containing open channels as topological node-line semimetals with Dirac nodal points forming a nodal loop in the $k_z$=0 plane of Brillouin zone. Interestingly, their nodal loops protected by inversion and time-reversal symmetries are robust against SU(2) symmetry breaking due to very weak spin-orbit coupling of Si. When the nodal lines are projected onto the (001) surface, flat surface bands can be observed because of the nontrivial topology of the bulk band structures. Our discoveries extend the topological physics to the three-dimensional Si materials, highlighting the possibility to realize low-cost, nontoxic and semiconductor-compatible Si-based electronics with topological quantum states.

[*] Corresponding author. Email: zhaojj@dlut.edu.cn

Silicon is the backbone of modern microelectronics due to its natural abundance, nontoxicity, propensity for doping, high-temperature stability, and native oxide passivation layer [1,2]. At ambient conditions, the ground state of Si crystal adopts the cubic diamond structure (*d*-Si) with $sp^3$ hybridization, which is an indirect-bandgap (1.17 eV) semiconductor. Moreover, the tetrahedral bonding character of Si atoms leads to an exceptionally complex energy landscape, giving rise to numerous metastable allotropes with small energy differences from the *d*-Si phase. By compressing or decompressing the known phases, many compact Si allotropes [3-6] have been discovered in laboratory. In addition, more and more low-density phases of Si have also been proposed and even synthesized [1,7-15]. These allotropes with different arrangements of Si atoms possess diverse electronic band structures, offering an effective way toward the desired physical properties for Si-based electronics. For instance, a number of direct- or quasidirect-bandgap semiconducting allotropes of Si [11-14] have been predicted by means of *ab initio* calculations, which could be utilized in the thin-film photovoltaic devices due to their modest bandgap and advanced optoelectronic properties beyond those of *d*-Si. Motivated by those theoretical studies, a recent experiment using a high-pressure precursor method reported the first synthesis of Si allotrope with quasidirect bandgap (1.3 eV), namely $Cmcm$-$Si_{24}$ [15]. On the other hand, Sung *et al.* proposed the first metallic phase ($P/6m$-$Si_6$) of Si in open framework, featuring as a superconductor with the critical temperature of ~12 K [16]. Notably, among the known metastable allotropes, BC8-Si was once predicted to be a semimetal [17,18], whereas recent experiment confirmed it is a narrow direct gap semiconductor [19]. Then a question arises naturally: is it possible to obtain a semimetal in the crystal structures of pure Si, especially a topological semimetal (TSM)?

Ignited by the successful theoretical design [20] and experimental observation [21,22] of Weyl semimetals (WSMs), the study of TSMs has attracted significant attention recently [23-27]. As new topological states of matter, TSMs are characterized by the non-accidental crossing between the conduction and valence bands near the Fermi level, and the linear energy dispersion as a function of momentum. The forming crossing-points (termed as nodes or nodal points) are either discrete or continuous. In the discrete case, according to the degeneracy of nodes, TSMs can be classified into three catalogs: (i) two-fold degenerate WSMs [20-23], (ii) triply degenerate nodal points semimetals (TDNPSMs) [28-31], and (iii) four-fold degenerate Dirac semimetals (DSMs) [32-35]. For the continuous situations, the nodal points form a periodically continuous line running across the Brillouin zone (BZ) or a closed loop inside the BZ; thus the corresponding systems are called topological node-line semimetals (TNLSMs). Under specific symmetry operations, quantum phase transition may occur between these TSM states. Importantly, their nontrivial topology encoded in the bulk band structure gives rise to remarkable physical properties, such as high carrier mobility, novel surface sates (e.g., Fermi arcs and drumhead-like flat bands), and Chiral anomaly [23-26]. In this regard, it is crucial to search for superb materials to realize the aforesaid TSM states for practical applications. After years of efforts, many fascinating TSM materials have been proposed and even confirmed experimentally [23-27]. However, most of them not only contain heavy elements that are often expensive or toxic, but also are hardly compatible with the current Si-based semiconductor industry.

In this letter, we show by first-principles calculations that it is indeed possible to obtain the desired TSM in the all-silicon solids. Upon screening 230 candidate structures in zeolite framework, we identify two open-framework Si allotropes termed as AHT-$Si_{24}$ and VFI-$Si_{36}$, belonging to TSMs. With all $sp^3$ bonding, these two structures show satisfactory stability, prevailing many other metastable Si allotropes. Both of them exhibit a novel TNLSM state with Dirac nodal points forming a closed loop within the $k_z = 0$ plane of BZ, and the flat bands around the Fermi level projected on the (001) surfaces. Distinct from most of the known TNLSMs formed by heavy elements, the symmetry-protected nodal lines in these two phases are robust against the breaking of SU(2) spin rotation symmetry at temperature higher than 14 K because of the weak spin-orbit coupling (SOC) of Si element.

The pioneering synthesis of *Cmcm*-$Si_{24}$ [15] has highlighted the possibility of unveiling new Si allotrope with zeolite structure since that *Cmcm*-$Si_{24}$ shares the same topology with the zeolite type code CAS in the database of International Zelolite Association (IZA) [36]. Interestingly, full examination for the lattice structures of the known Si allotropes reveals that the prominent Si clathrates [7,37] *i.e.*, $Si_{46}$ and $Si_{136}$, also correspond to the zeolite frameworks of MTN and MEP, respectively. Likewise, the recently predicted phase hP12-Si from particle swarm optimization algorithm search [12], has identical framework of the zeolite CAN. Considering these facts and the four-fold-coordinated character of all zeolite frameworks, it is reasonable to search the possible TSM allotropes by screening the Si crystals constructed from the IZA database (see Fig. S1 of the Supplemental Materials for a flow of screening [38]).

The atomic and electronic structures of these 230 all-Si candidate structures from the IZA database were computed using density functional theory as implemented in the Vienna *ab initio* simulation package [39]. The Perdew-Burke-Ernzerhof functional [40] within the generalized gradient approximation (GGA) was employed to treat the exchange-correlation interactions. The hybrid functional HSE06 [41] was further employed to validate the PBE band structures for the semimetallic Si phases. The core-valence interactions were described by the PAW method [42] with a kinetic energy cutoff of 500 eV. All atoms were fully relaxed until total energy and atomic force were less than $10^{-6}$ eV and $10^{-3}$ eV/Å, respectively. Monkhorst-Pack *k*-point meshes with a uniform density of $2\pi \times 0.01$ Å$^{-1}$ were generated to sample the BZ. To explore the nontrivial topological properties of the TSM phases, the maximally localized Wannier functions [43] were constructed to obtain the tight-binding Hamiltonian; then the surface states were calculated by the iterative Green's function method with the WannierTools package [44].

Upon relaxation, all these hypothetic zeolite-like Si framework structures are well preserved. With such abundant Si crystal lattices, a rich variety of electronic properties is observed from the calculated PBE band structures, ranging from direct and indirect semiconductors, metals, to semimetals. The key data are listed in Table SI of the Supplemental Materials [38]. For the plausible semimetallic systems, more accurate HSE06 calculations were conducted to validate the PBE band structures. Finally, two zeolite-like Si structures, *i.e.*, AHT-$Si_{24}$ and VFI-$Si_{36}$ with linear bands crossing near the Fermi level, are confirmed as semimetals. Hereafter, we will focus on AHT-$Si_{24}$ as a representative.

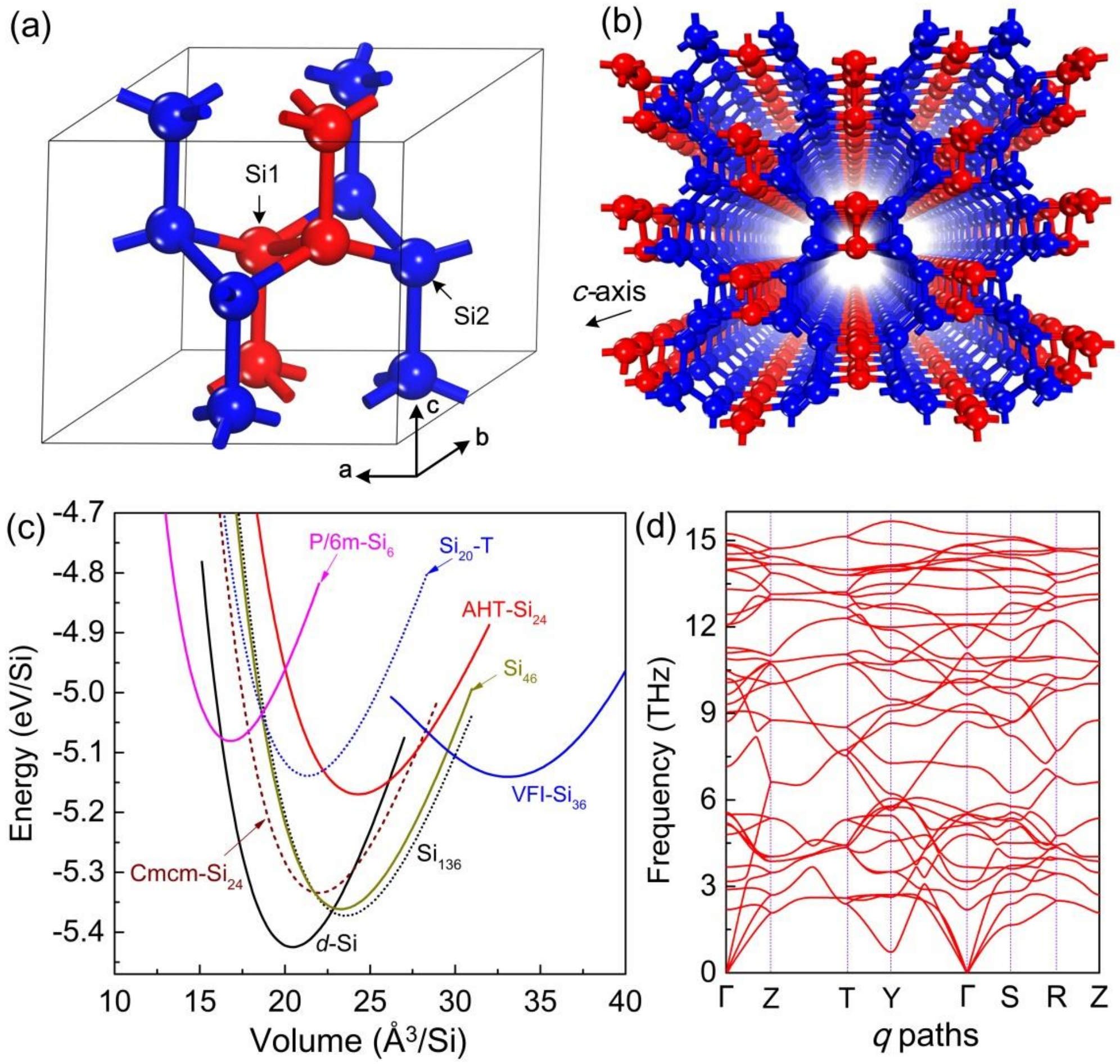

FIG. 1. (a) The primitive cell of AHT-$Si_{24}$ with *Cmcm* ($D_{2h}^{17}$) space group consisting of two kinds of crystallographically inequivalent atoms: Si1 (red) and Si2 (blue) at 8*f* (0, 0.3504, 0.9361) and 16*h* (0.6829, 0.1504, 0.5639) Wyckoff positions, respectively. The lattice constants of AHT-$Si_{24}$: $a$ = 12.18 Å, $b$ = 7.52 Å, $c$ = 6.37 Å. (b) A perspective view from crystallographic *c*-axis for the structure of AHT-$Si_{24}$. (c) The cohesive energy as a function of volume per Si atom for AHT-$S_{24}$ and VFI-$Si_{36}$ in comparison with some known Si allotropes. (d) Phonon dispersion of AHT-$Si_{24}$ confirming its dynamic stability.

Fig. 1(a) shows the crystal structure of AHT-$Si_{24}$ which has two kinds of inequivalent Si atoms, marked as Si1 and Si2. All the atoms are four-fold coordinated, forming $sp^3$-like hybridization (each Si1 atom bonds with two Si1 and two Si2 atoms, while every Si2 is connected to one Si1 and three Si2 atoms). The average bond angle is 107.58°, rather close to the perfect tetrahedral angle of 109.47° in *d*-Si phase. Similar to *Cmcm*-$Si_{24}$ [15], AHT-$Si_{24}$ is featured by periodic open channels constructed by ten-membered rings along the crystallographic *c*-axis [see Fig. 1(b)]. This suggests that the AHT-$Si_{24}$ phase could also be synthesized by the high-pressure precursor method [15] like *Cmcm*-$Si_{24}$. If a precursor Na-Si system, e.g., AHT-$Na_4Si_{24}$ (see Fig. S3 [38]), is produced and retains its stability at ambient condition, the guest Na atoms could be removed from the open channels by a degassing process [15,16,45].

The calculated cohesive energy of AHT-$Si_{24}$ is 5.17 eV, which is less than that of *d*-Si by 0.25 eV/atom [Fig. 1(c)], due to the ring strain in the open channels. In spite of this, it is more stable than the open-framework allotropes $P/6m$-$Si_6$ (5.07 eV) [16] and $Si_{20}$-T (5.14 eV) [11], and other synthesized compact allotropes of Si, including $\beta$-Sn, *Imma*-Si, SH-Si, *Cmca*-Si and Si-VII, and Si-X (see Fig. S4 [38]) at zero pressure. The dynamic stability of AHT-$Si_{24}$ has been confirmed by phonon dispersion calculation, as shown in Fig. 2(d). Moreover, its independent elastic constants $C_{11}$, $C_{12}$, $C_{13}$, $C_{22}$, $C_{23}$, $C_{33}$, $C_{44}$, $C_{55}$, and $C_{66}$ are 134, 36, 29, 96, 24, 162, 27, 31, and 36 GPa, respectively, satisfying the mechanical stability criteria [*i.e.*, $C_{ii} > 0$ ($i$=1, 2, ..., 6), $C_{11} + C_{22} + C_{33} + 2\,(C_{12} + C_{13} + C_{23}) > 0$, $C_{11} + C_{22} - 2C_{12} > 0$, $C_{11} + C_{33} - 2C_{13} > 0$, $C_{22} +$

$C_{33} - 2C_{23} > 0$] for an orthorhombic crystal [46].

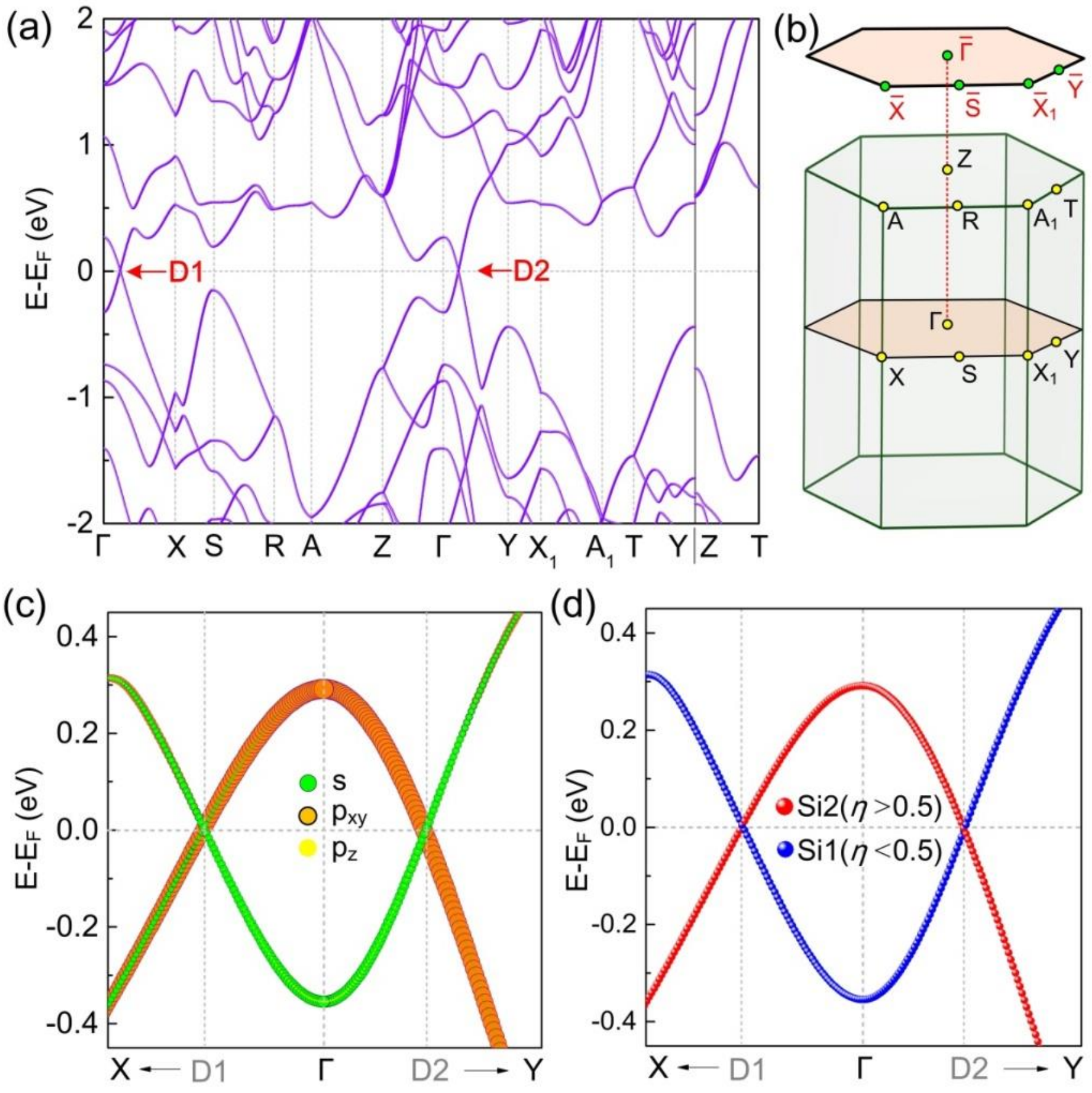

FIG. 2. (a) Bulk electronic band structure of AHT-$Si_{24}$. (b) The corresponding bulk and (001)-surface BZ together with the high-symmetry *k*-points. (c) Orbital decomposed linear crossing bands around the nodal points. (d) Colorful band structure near the nodal points weighted by the contribution ratio, *i.e.*, $\eta = W_{Si2} / (W_{Si1} + W_{Si2})$, where W denotes the weight of the corresponding atom.

The electronic band structure of AHT-$Si_{24}$ is presented in Fig. 2(a). Along two perpendicular paths, namely Γ→X and Γ→Y, the highest valence band (HVB) and lowest conduction band (LCB) exhibit Dirac linear dispersions in the vicinity of the Fermi energy, and cross at asymmetric Dirac nodal points D1 ( $\lambda_1 \sec^2\theta/4$, $\lambda_1 \sec^2\theta/4$,

0) and D2 ( $\lambda_2/2$ , $\lambda_2/2$ , 0) on the Fermi level. Here, $\lambda_1$ = 0.241 and $\lambda_2$ = 0.228 are the length ratios between Γ-D1 and Γ-X, and between Γ-D2 and Γ-Y, respectively; $\theta$ = 31.69 ° is half of the angle between $\boldsymbol{k}_\mathrm{x}$ and $\boldsymbol{k}_\mathrm{y}$ vectors in the BZ. Note that the linear crossing also exists in the HSE06 band structure (Fig. S5 [38]). To further explore the origin of band crossings in AHT-$Si_{24}$, we calculated the orbital and atom decomposed band structures near D1 and D2. As shown in Fig. 2(c) and 2(d), the crossing bands are mainly originated from the $p_{xy}$ states of Si2 atoms termed as $\left|\mathrm{Si2}, p_{xy}\right\rangle$, and the $s$ and $p_{xy}$ hybridized states of Si1 atoms denoted as $\left|\mathrm{Si1}, s{+}p_{xy}\right\rangle$. Clearly, there exists exchange of energy ordering between $\left|\mathrm{Si2}, p_{xy}\right\rangle$ and $\left|\mathrm{Si1}, s{+}p_{xy}\right\rangle$, resulting in a band crossing. Importantly, such band inversion is one of the key features for topological insulators [47,48] as well as TNLSMs [24,49].

To further reveal the pattern of the crossing nodal points in the entire BZ, we calculated the Fermi surface and 3D band structure of AHT-$Si_{24}$. The results show that HVB and LCB cross with each other along a closed loop [Fig. 3(a)], which locates on the $k_\mathrm{z}$ = 0 plane of BZ due to the mirror reflection symmetry. Hence, the AHT-$Si_{24}$ is a new member of topological node-line semimetals. A closer examination shows that the nodal loop is a rounded rectangle with the long edge parallel to ΓX line and the short one parallel to ΓY line [Fig. 3(b)], which is a consequence of the orthorhombic crystalline symmetry. Naturally, one may wonder whether the low-energy Dirac Fermions along the two perpendicular directions exhibit anisotropic transport behavior? To explore this, the Fermi velocities, denoted as $v_\mathrm{F}$, have been evaluated by a linear fittings [$\hbar v_\mathrm{F} \approx \mathrm{d}E(\boldsymbol{k})/\mathrm{d}\boldsymbol{k}$] for the linear crossing bands. Along the D1→X direction, *i.e.*, the long side of the rounded rectangle, the Fermi velocities of the hole and electron carriers are $4.86\times10^5$ m/s and $5.44\times10^5$ m/s, respectively. Along the D2→Y direction, *i.e.*, the short side of the rounded rectangle, the Fermi velocities of the hole and electron carriers are $7.76\times10^5$ m/s and $6.87\times10^5$ m/s, respectively. In other words, the carrier transport in AHT-$Si_{24}$ is anisotropic. Remarkably, these velocities are comparable with that of graphene ($9.42\times10^5$ m/s [50]) and even larger than that of silicene ($5.31\times10^5$ m/s [51]).

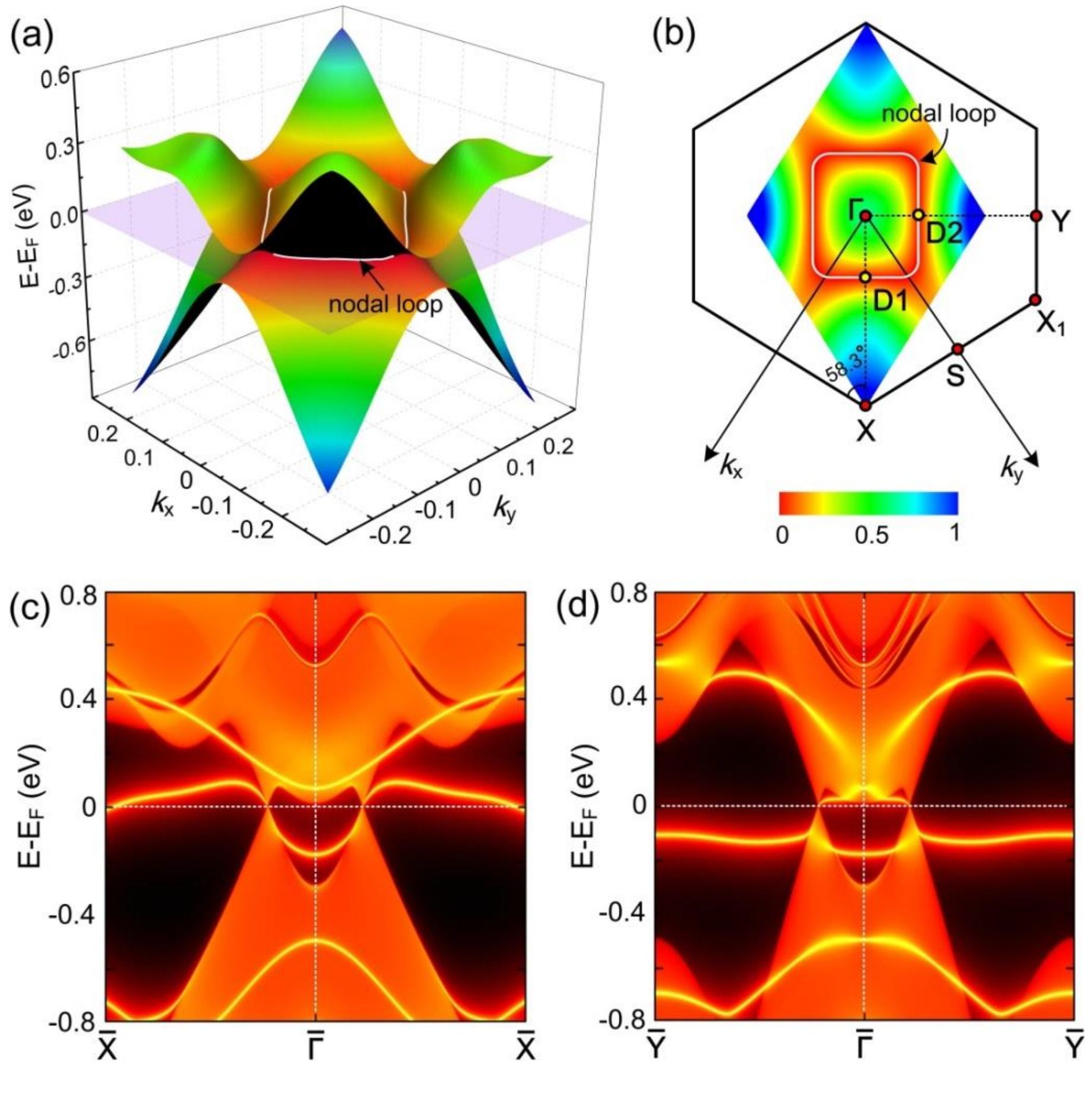

FIG. 3. (a) 3D band structure of the HVB and LCB for AHT-$Si_{24}$. (b) The colorized contour plot of the energy band gap between HVB and LCB in the $k_z$=0 plane. The (001)-surface states along (c) $\overline{\Gamma}-\overline{X}$ and (d) $\overline{\Gamma}-\overline{Y}$ directions within the $\overline{\Gamma}-\overline{X}-\overline{S}-\overline{X}_1-\overline{Y}$ plane [see Fig. 2(b)].

According to the classification of TNLSMs [25], we infer that the AHT-$Si_{24}$ belongs to type-B because its closed nodal line is protected by space inversion (*P*), time-reversal (*T*), and SU(2) spin-rotation symmetries. Specifically, if we enclose the nodal line with a ring, the independent $Z_2$ index, termed as $\zeta_1$, should be topologically invariant. Here, the explicit expression of $\zeta_1$ is the Berry phase on the ring that interlocks with the closed nodal line [25]: $(-1)^{\zeta_1} = \oint \mathrm{d}k\boldsymbol{A}(\boldsymbol{k}) \cdot \mathrm{d}\boldsymbol{k}$ , where $\boldsymbol{A}(\boldsymbol{k}) = -i \sum_{n\in\mathrm{occ}} \langle u_n(\boldsymbol{k}) | \partial_{\boldsymbol{k}} | u_n(\boldsymbol{k}) \rangle$ is the Abelian Berry connection. For AHT-$Si_{24}$, $\zeta_1$ is equal to 1, implying that the rounded-rectangle nodal line is topologically protected by the mentioned *P*, *T* and SU(2) symmetries. If these symmetries are partially broken, what would be the fate of the nodal line? To investigate the possible topological phase transition, we carried out first-principles calculations of AHT-$Si_{24}$ with the breaking of SU(2) symmetry (or by considering SOC). As known, in the presence of SOC, the band repulsion between opposite spins is usually non-negligible, making the closed nodal line unstable [24-26]. For AHT-$Si_{24}$, however, like the reported TNLSM materials formed by light elements [49,52-55], the computed SOC splitting is remarkably small (around 0.7 meV and 0.4 meV along Γ-X and Γ-Y lines, respectively), which can be ignored completely at temperature higher than 9 K. In this sense, the topological nodal line in AHT-$Si_{24}$ is rather robust against SOC under ambient temperature. Thus the needed quantum phase transitions should resort to the symmetry breakings of *P*, *T* or both of them [24,25].

For a TNLSM, another key feature is the existence of flat surface bands [24], which provides the possibility for experimental detection by angle-resolved photoemission measurement [56]. The (001)-surface states of AHT-$Si_{24}$ along $\bar{\Gamma}-\bar{X}$ and $\bar{\Gamma}-\bar{Y}$ directions are presented in Figs. 3(c) and 3(d), respectively. Around the Fermi level, two types of nearly flat bands can be observed, which nestle either inside or outside of the nodal line — rounded rectangle. In the $\bar{\Gamma}-\bar{X}$ direction, the outside flat band lies above the Fermi level; in contrast, the outside flat band locates below the Fermi level in the $\bar{\Gamma}-\bar{Y}$ direction. This means that the surface carrier transport based on these flat bands of AHT-$Si_{24}$ are anisotropic, similar to that of the Dirac fermions in the bulk states. For practical application, such flat surface bands could induce interesting physical effects, *e.g.*, two-dimensional anisotropic surface superconductivity with high critical temperature [57,58].

We have also performed systematical calculations on VFI-$Si_{36}$ phase, and the detailed results are shown in Figs. S6-S8 in the Supplemental Materials [38]. Essentially, we can draw the same conclusions as those for AHT-$Si_{24}$. As for VFI-$Si_{36}$, the HVB and LCB linearly cross at the asymmetric points D1 along Γ-M and D2 along Γ-K, which are caused by the band inversion between $\left|\mathrm{Si1{+}Si2}, p_{xy}\right\rangle$ and $\left|\mathrm{Si2}, s{+}p_{xy}\right\rangle$ states. Due to the $P6_3/mcm$ ($D_{6h}^3$)symmetry, its nodal line winds into a closed rounded-hexagon in the $k_z$=0 plane. The computed SOC splitting is about 1.2 meV and 0.6 meV along Γ-K and Γ-M paths, respectively, which can be ignored at temperature higher than 14 K. Moreover, the projected flat surface bands can be found around the Fermi level as well.

To summarize, using first-principles calculations we have verified two Si allotropes with zeolite framework as topological semimetals, namely, AHT-$Si_{24}$ and VFI-$Si_{36}$. Their structural stabilities were confirmed from cohesive energy, phonon dispersion and elastic constants. These two allotropes hold open channels along certain crystallographic axis, similar to the structural characteristics of the recently synthesized *Cmcm*-$Si_{24}$; thus they can be possibly fabricated by the high-pressure precursor method. The calculated electronic band structures uncover that both AHT-$Si_{24}$ and VFI-$Si_{36}$ exhibit a novel TNLSM state, in which the Dirac nodal points with linear dispersion form a closed ring within the $k_z$=0 plane of BZ. More importantly, the *P* and *T* symmetry-protected nodal lines are rather robust in the presence of SOC under room temperature. Furthermore the flat surface bands have been observed on their (001) surface, giving rise to exotic transport behavior (e.g., high-temperature anisotropic surface superconductivity) and other strong correlation physical properties [59-61]. Our prediction of TNLSM Si phases may pave a way to realize the all-silicon topological devices with low-cost, nontoxic and semiconductor-compatible features.

## ACKNOWLEDGMENTS

We acknowledge useful discussions with Prof. S.-W. Gao. This work is supported by the National Natural Science Foundation of China (11604165, 11574040) and Natural Science Foundation of Inner Mongolia (2016BS0104). The Supercomputing Center of Dalian University of Technology is acknowledged for providing computing resources.

Supporting Information for

# All-Silicon Topological Semimetals with Closed Nodal Line

Zhifeng Liu,[1,2] Hongli Xin,[1] Li Fu,[1] Yingqiao Liu,[3] Tielei Song,[1] Xin Cui,[1]
Guojun Zhao,[1] Jijun Zhao[2,3*]

[1]*School of Physical Science and Technology, Inner Mongolia University, Hohhot 010021, China*

[2]*Beijing Computational Science Research Center, Beijing 100094, China*

[3]*Key Laboratory of Materials Modification by Laser, Ion and Electron Beams* (*Dalian University of Technology*), *Ministry of Education, Dalian 116024, China*

* Corresponding author. Email: zhaojj@dlut.edu.cn

## I. The Flow of Screening

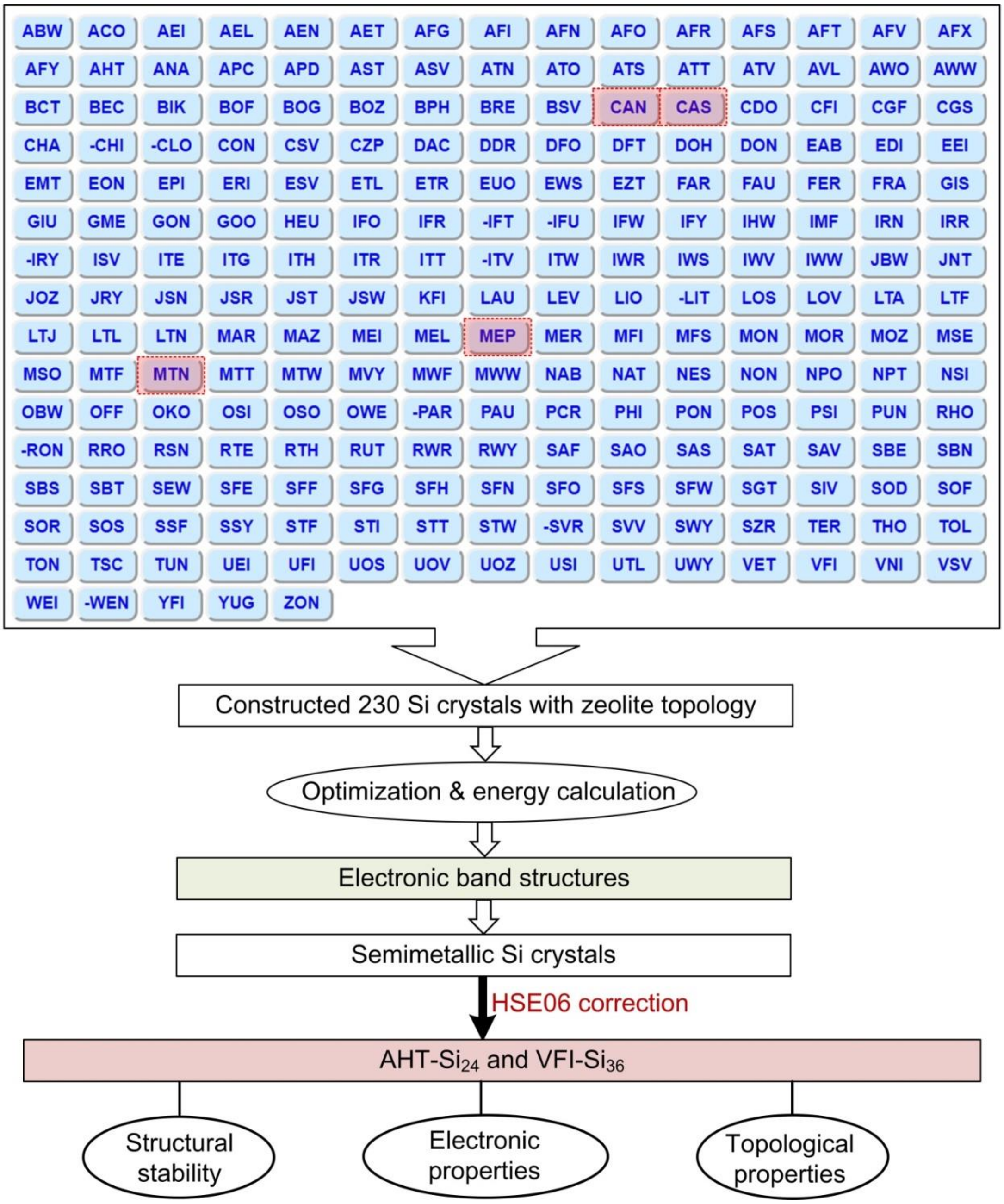

FIG. S1. The flow of screening for the semimetallic Si crystals with zeolite topology. On the top panel, we present all the considered 230 zeolite framework codes. For more details about their topology, please see http://www.iza-structure.org/databases/. The shaded codes mean that their corresponding Si allotropes have been discovered in previous reports (*i.e.*, hp12-Si with CAN framework, *Cmcm*-$Si_{24}$ with CAS, Clathrate $Si_{46}$ with MTN and $Si_{136}$ with MEP).

## II. The Key Data for All of the Considered Zeolite Si Crystals

**Table SI** The space group (*SG*), lattice constants (*a*, *b* and *c*, Å), mass density ($\rho$, *g cm*$^{-3}$), equilibrium volume per atom ($V_0$, Å$^3$), cohesive energy ($E_c$, eV/atom), band gap ($E_g$, eV) and the electronic properties [*P*, metal (M), semimetal (SM), direct bandgap semiconductor (DS), indirect semiconductor (IS)] for the zeolite Si allotropies optimized at PBE level of theory. The energy as a function of volume for all of these Si-allotropes is presented in Fig. S2.

| Si-phases | *SG* | *a* | *b* | *c* | $\rho$ | $V_0$ | $E_c$ | $E_g$ | *P* |
|---|---|---|---|---|---|---|---|---|---|
| ABW-$Si_8$ | *Imma* | 3.90 | 6.53 | 7.53 | 1.95 | 23.97 | 5.22 | 0.00 | M |
| ACO-$Si_{16}$ | *Im*-3*m* | 7.49 | 7.49 | 7.49 | 1.77 | 26.28 | 5.10 | 0.10 | DS |
| AEI-$Si_{48}$ | *Cmcm* | 9.66 | 10.50 | 14.11 | 1.56 | 29.82 | 5.07 | 0.58 | IS |
| AEL-$Si_{40}$ | *Imma* | 6.38 | 10.37 | 13.98 | 2.02 | 23.12 | 5.29 | 0.61 | DS |
| AEN-$Si_{48}$ | *Cmce* | 10.86 | 13.67 | 7.29 | 2.07 | 22.55 | 5.20 | 0.00 | M |
| AET-$Si_{72}$ | *Cmcm* | 11.48 | 25.31 | 6.37 | 1.81 | 25.71 | 5.23 | 0.00 | M |
| AFG-$Si_{48}$ | *P6_3/mmc* | 9.45 | 9.45 | 15.67 | 1.85 | 25.25 | 5.20 | 0.00 | SM |
| AFI-$Si_{24}$ | *P6/mcc* | 10.51 | 10.51 | 6.38 | 1.83 | 25.44 | 5.26 | 0.35 | IS |
| AFN-$Si_{32}$ | *C2/m* | 10.96 | 10.18 | 8.20 | 1.74 | 26.79 | 5.07 | 0.58 | IS |
| AFO-$Si_{40}$ | *Cmcm* | 7.43 | 19.56 | 6.37 | 2.01 | 23.17 | 5.28 | 0.77 | IS |
| AFR-$Si_{32}$ | *Pmmn* | 5.39 | 10.53 | 17.12 | 1.54 | 30.32 | 5.09 | 0.79 | IS |
| AFS-$Si_{56}$ | *P6_3/mcm* | 10.18 | 10.18 | 19.98 | 1.46 | 32.04 | 5.05 | 0.61 | IS |
| AFT-$Si_{72}$ | *P6_3/mmc* | 10.54 | 10.54 | 22.26 | 1.57 | 29.75 | 5.07 | 0.77 | IS |
| AFV-$Si_{30}$ | *P*-3*m*1 | 10.14 | 10.14 | 9.66 | 1.63 | 28.68 | 5.08 | 0.65 | IS |
| AFX-$Si_{48}$ | *P6_3/mmc* | 10.53 | 10.53 | 14.86 | 1.57 | 29.74 | 5.07 | 0.66 | DS |
| AFY-$Si_{16}$ | *P-31m* | 9.75 | 9.75 | 6.75 | 1.34 | 34.78 | 4.94 | 0.28 | IS |
| AHT-$Si_{24}$ | *Cmcm* | 12.18 | 7.52 | 6.37 | 1.92 | 24.33 | 5.17 | 0.00 | SM |
| ANA-$Si_{48}$ | *Ia*-3*d* | 10.35 | 10.35 | 10.35 | 2.02 | 23.07 | 5.15 | 0.94 | DS |
| APC-$Si_{32}$ | *Cmce* | 6.94 | 15.08 | 7.75 | 1.84 | 25.35 | 5.14 | 0.52 | IS |
| APD-$Si_{32}$ | *Cmce* | 6.48 | 15.35 | 7.63 | 1.97 | 23.69 | 5.24 | 0.68 | IS |
| AST-$Si_{40}$ | *Fm*-3*m* | 10.25 | 10.25 | 10.25 | 1.73 | 26.92 | 5.13 | 0.11 | IS |
| ASV-$Si_{20}$ | *P4/mcc* | 6.47 | 6.47 | 11.24 | 1.98 | 23.53 | 4.96 | 0.00 | M |
| ATN-$Si_{16}$ | *I4/mmm* | 3.88 | 14.22 | 7.37 | 1.90 | 24.56 | 5.19 | 0.00 | M |
| ATO-$Si_{36}$ | *R*-3*m* | 15.98 | 15.98 | 3.82 | 1.99 | 23.47 | 5.23 | 0.47 | IS |

| **Si-phases** | ***SG*** | ***a*** | ***b*** | ***c*** | ***ρ*** | $V_0$ | $E_c$ | $E_g$ | ***P*** |
|---|---|---|---|---|---|---|---|---|---|
| ATS-$Si_{24}$ | *Cmcm* | 10.29 | 16.50 | 3.87 | 1.70 | 27.36 | 5.16 | 0.00 | SM |
| ATT-$Si_{12}$ | *Pmma* | 5.88 | 7.10 | 7.46 | 1.80 | 25.98 | 5.10 | 0.28 | IS |
| ATV-$Si_{24}$ | *Cmme* | 6.38 | 11.62 | 7.06 | 2.14 | 21.79 | 5.32 | 0.74 | IS |
| AVL-$Si_{42}$ | *P*-3*m*1 | 10.24 | 10.24 | 13.47 | 1.60 | 29.12 | 5.07 | 0.69 | IS |
| AWO-$Si_{48}$ | *Cmce* | 6.94 | 11.86 | 14.44 | 1.88 | 24.76 | 5.13 | 0.56 | IS |
| AWW-$Si_{24}$ | *P4/nmm* | 10.55 | 10.55 | 5.80 | 1.73 | 26.88 | 5.13 | 0.48 | DS |
| BCT-$Si_{8}$ | *I4/mmm* | 6.68 | 6.68 | 3.87 | 2.16 | 21.60 | 5.32 | 0.26 | IS |
| BEC-$Si_{32}$ | *P4_2/mmc* | 9.74 | 9.74 | 10.11 | 1.56 | 29.99 | 5.13 | 0.73 | IS |
| BIK-$Si_{12}$ | *Cmcm* | 5.74 | 11.97 | 3.85 | 2.12 | 22.02 | 5.34 | 0.43 | IS |
| BOF-$Si_{24}$ | *Pnma* | 5.71 | 9.83 | 10.42 | 1.92 | 24.35 | 5.19 | 0.80 | DS |
| BOG-$Si_{96}$ | *Imma* | 9.82 | 15.23 | 18.45 | 1.62 | 28.74 | 5.17 | 0.93 | IS |
| BOZ-$Si_{92}$ | *Cmcm* | 11.24 | 29.21 | 10.96 | 1.19 | 39.12 | 4.74 | 0.18 | DS |
| BPH-$Si_{28}$ | *P*-62*m* | 10.18 | 10.18 | 10.02 | 1.45 | 32.11 | 5.05 | 0.50 | DS |
| BRE-$Si_{16}$ | *P2_1/m* | 5.17 | 13.06 | 5.91 | 1.87 | 24.94 | 5.16 | 0.94 | IS |
| BSV-$Si_{96}$ | *Ia*-3*d* | 13.98 | 13.98 | 13.98 | 1.64 | 28.45 | 4.79 | 0.04 | IS |
| CAN-$Si_{12}$ | *P6_3/mmc* | 9.45 | 9.45 | 3.91 | 1.85 | 25.17 | 5.20 | 0.00 | SM |
| CAS-$Si_{24}$ | *Cmcm* | 3.86 | 10.74 | 12.73 | 2.12 | 22.00 | 5.33 | 0.51 | IS |
| CDO-$Si_{36}$ | *Cmcm* | 5.73 | 14.00 | 10.70 | 1.96 | 23.85 | 5.25 | 0.36 | IS |
| CFI-$Si_{32}$ | *Imma* | 3.98 | 10.57 | 19.44 | 1.83 | 25.53 | 5.27 | 0.35 | IS |
| CGF-$Si_{36}$ | *C2/m* | 12.08 | 13.59 | 5.71 | 1.80 | 25.88 | 5.09 | 0.46 | IS |
| CGS-$Si_{32}$ | *Pnma* | 6.62 | 11.21 | 12.32 | 1.63 | 28.58 | 5.02 | 0.80 | IS |
| CHA-$Si_{36}$ | *R*-3*m* | 10.53 | 10.53 | 11.12 | 1.57 | 29.70 | 5.07 | 0.73 | IS |
| CHI-$Si_{28}$ | *Pbcn* | 3.94 | 23.08 | 6.97 | 2.06 | 22.62 | 5.06 | 0.00 | M |
| CLO-$Si_{192}$ | *Pm-3m* | 20.18 | 20.18 | 20.18 | 1.09 | 42.83 | 4.95 | 0.39 | DS |
| CON-$Si_{56}$ | *C2/m* | 17.27 | 10.55 | 9.54 | 1.61 | 29.04 | 5.16 | 0.87 | IS |
| CSV-$Si_{20}$ | *P*-1 | 7.25 | 8.62 | 10.00 | 1.62 | 28.84 | 5.11 | 0.69 | IS |
| CZP-$Si_{24}$ | *P6_122* | 7.44 | 7.44 | 12.07 | 1.93 | 24.14 | 4.85 | 0.00 | M |
| DAC-$Si_{24}$ | *C2/m* | 14.25 | 5.76 | 7.57 | 1.90 | 24.57 | 5.24 | 0.45 | IS |
| DDR-$Si_{120}$ | *R*-3*m* | 10.50 | 10.50 | 31.83 | 1.84 | 25.31 | 5.29 | 1.20 | IS |
| DFO-$Si_{132}$ | *P6/mmm* | 17.20 | 17.20 | 16.24 | 1.48 | 31.54 | 5.06 | 0.47 | IS |
| DFT-$Si_{8}$ | *P4_2/mmc* | 5.46 | 5.46 | 6.59 | 1.89 | 24.62 | 5.18 | 0.59 | IS |
| DOH-$Si_{34}$ | *P6/mmm* | 10.51 | 10.51 | 8.56 | 1.94 | 24.09 | 5.34 | 1.14 | IS |
| DON-$Si_{64}$ | *Cmcm* | 14.69 | 17.63 | 6.54 | 1.77 | 26.36 | 5.24 | 0.53 | IS |

| Si-phases | *SG* | *a* | *b* | *c* | *ρ* | $V_0$ | $E_c$ | $E_g$ | *P* |
|---|---|---|---|---|---|---|---|---|---|
| EAB-$Si_{36}$ | *P6_3/mmc* | 9.95 | 9.95 | 11.76 | 1.67 | 27.98 | 5.09 | 0.72 | IS |
| EDI-$Si_5$ | *P*-4*m*2 | 5.25 | 5.25 | 4.95 | 1.71 | 27.29 | 5.12 | 0.96 | DS |
| EEI-$Si_{50}$ | *Fmmm* | 9.97 | 9.97 | 14.40 | 1.95 | 23.96 | 5.31 | 0.89 | IS |
| EMT-$Si_{96}$ | *P6_3/mmc* | 13.43 | 13.43 | 22.01 | 1.30 | 35.82 | 5.00 | 0.27 | IS |
| EON-$Si_{60}$ | *Pmmn* | 5.78 | 14.00 | 19.75 | 1.75 | 26.65 | 5.13 | 0.59 | DS |
| EPI-$Si_{24}$ | *C2/m* | 6.94 | 13.58 | 6.81 | 1.90 | 24.51 | 5.23 | 0.58 | IS |
| ERI-$Si_{36}$ | *P6_3/mmc* | 9.93 | 9.93 | 11.70 | 1.68 | 27.75 | 5.09 | 0.03 | IS |
| ESV-$Si_{48}$ | *Pnma* | 7.51 | 9.43 | 17.60 | 1.80 | 25.96 | 5.21 | 1.19 | IS |
| ETL-$Si_{72}$ | *Cmcm* | 5.88 | 22.03 | 13.35 | 1.94 | 24.02 | 5.22 | 0.54 | IS |
| ETR-$Si_{48}$ | *P6_3mc* | 16.28 | 16.28 | 6.64 | 1.47 | 31.64 | 5.00 | 0.77 | DS |
| EUO-$Si_{112}$ | *Cmme* | 10.61 | 16.99 | 15.50 | 1.87 | 24.94 | 5.28 | 0.95 | DS |
| EWS-$Si_{96}$ | *Cmce* | 12.21 | 14.62 | 14.23 | 1.77 | 26.39 | 5.16 | 0.60 | IS |
| EZT-$Si_{48}$ | *Imma* | 7.93 | 9.82 | 16.84 | 1.71 | 27.33 | 5.09 | 0.00 | M |
| FAR-$Si_{84}$ | *P6_3/mmc* | 9.49 | 9.49 | 27.25 | 1.84 | 25.33 | 5.20 | 0.06 | DS |
| FAU-$Si_{192}$ | *F*d-3*m* | 19.00 | 19.00 | 19.00 | 1.31 | 35.71 | 5.00 | 0.28 | IS |
| FER-$Si_{36}$ | *Immm* | 5.75 | 10.78 | 14.11 | 1.92 | 24.27 | 5.25 | 0.48 | DS |
| FRA-$Si_{60}$ | *P*-3*m*1 | 9.51 | 9.51 | 19.45 | 1.84 | 25.38 | 5.20 | 0.16 | IS |
| GIS-$Si_{16}$ | *I4_1/amd* | 7.51 | 7.51 | 7.76 | 1.71 | 27.33 | 5.07 | 0.46 | IS |
| GIU-$Si_{96}$ | *P6_3/mmc* | 9.48 | 9.48 | 31.16 | 1.85 | 25.27 | 5.20 | 0.00 | SM |
| GME-$Si_{24}$ | *P6_3/mmc* | 10.53 | 10.53 | 7.47 | 1.56 | 29.87 | 5.06 | 0.24 | DS |
| GON-$Si_{32}$ | *Cmmm* | 12.85 | 15.31 | 3.88 | 1.95 | 23.89 | 5.29 | 0.43 | IS |
| GOO-$Si_{32}$ | *C*222_1 | 6.68 | 9.02 | 13.22 | 1.87 | 24.95 | 5.09 | 0.72 | IS |
| HEU-$Si_{36}$ | *C2/m* | 13.10 | 13.45 | 5.67 | 1.82 | 25.60 | 5.18 | 0.64 | IS |
| IFO-$Si_{32}$ | *Pnnm* | 3.80 | 12.94 | 18.12 | 1.67 | 27.85 | 5.14 | 0.36 | IS |
| IFR-$Si_{32}$ | *C2/m* | 14.51 | 10.61 | 5.86 | 1.69 | 27.54 | 5.13 | 0.49 | IS |
| IFT-$Si_{152}$ | *Cmcm* | 14.75 | 17.37 | 22.79 | 1.21 | 38.50 | -5.01 | 0.00 | M |
| IFU-$Si_{64}$ | *C2/m* | 12.08 | 15.34 | 15.35 | 1.23 | 37.93 | 4.98 | 0.00 | M |
| IFW-$Si_{64}$ | *C2/m* | 16.47 | 13.91 | 9.61 | 1.66 | 28.06 | 5.17 | 0.54 | IS |
| IFY-$Si_{48}$ | *P4/mbm* | 11.87 | 11.87 | 9.29 | 1.71 | 27.30 | 5.07 | 0.45 | IS |
| IHW-$Si_{112}$ | *Cmce* | 10.68 | 18.37 | 13.82 | 1.93 | 24.21 | 5.27 | 0.92 | IS |
| IMF-$Si_{280}$ | *Cmcm* | 11.00 | 43.19 | 15.13 | 1.82 | 25.65 | 5.18 | 0.30 | IS |
| IRN-$Si_{92}$ | *Immm* | 12.78 | 13.89 | 15.07 | 1.61 | 29.05 | 5.10 | 0.59 | IS |
| IRR-$Si_{52}$ | *P6/mmm* | 15.10 | 15.10 | 10.95 | 1.12 | 41.71 | 5.01 | 0.00 | M |

| **Si-phases** | ***SG*** | ***a*** | ***b*** | ***c*** | ***ρ*** | $V_0$ | $E_c$ | $E_g$ | ***P*** |
|---|---|---|---|---|---|---|---|---|---|
| IRY-$Si_{76}$ | *P6_3/mmc* | 12.55 | 12.55 | 24.79 | 1.05 | 44.51 | 4.96 | 0.35 | IS |
| ISV-$Si_{64}$ | *P4_2/mmc* | 9.94 | 9.94 | 19.53 | 1.55 | 30.12 | 5.12 | 0.73 | IS |
| ITE-$Si_{64}$ | *Cmcm* | 7.55 | 16.04 | 14.92 | 1.65 | 28.22 | 5.15 | 0.74 | IS |
| ITG-$Si_{56}$ | *P2/m* | 9.96 | 9.67 | 16.17 | 1.69 | 27.62 | 5.18 | 1.04 | IS |
| ITH-$Si_{56}$ | *Amm*2 | 8.75 | 17.00 | 9.71 | 1.81 | 25.79 | 5.22 | 0.83 | IS |
| ITR-$Si_{112}$ | *Cmcm* | 8.75 | 17.00 | 19.40 | 1.81 | 25.77 | 5.21 | 0.86 | IS |
| ITT-$Si_{46}$ | *P6/mmm* | 15.01 | 15.01 | 8.68 | 1.27 | 36.81 | 5.08 | 0.62 | IS |
| ITV-$Si_{192}$ | *P4_132* | 20.39 | 20.39 | 20.39 | 1.06 | 44.18 | 4.92 | 0.00 | M |
| ITW-$Si_{24}$ | *C2/m* | 7.62 | 11.71 | 6.84 | 1.89 | 24.71 | 5.13 | 0.57 | IS |
| IWR-$Si_{56}$ | *Cmmm* | 10.54 | 16.10 | 9.64 | 1.60 | 29.21 | 5.14 | 0.56 | DS |
| IWS-$Si_{136}$ | *I4/mmm* | 20.62 | 20.62 | 9.96 | 1.50 | 31.14 | 5.11 | 0.41 | IS |
| IWV-$Si_{38}$ | *Fmmm* | 10.68 | 11.10 | 12.11 | 1.57 | 29.74 | 5.19 | 0.71 | IS |
| IWW-$Si_{112}$ | *Pbam* | 9.67 | 9.93 | 32.04 | 1.70 | 27.48 | 5.18 | 1.18 | IS |
| JBW-$Si_{6}$ | *Pmma* | 3.87 | 5.86 | 5.94 | 2.08 | 22.44 | 5.27 | 0.25 | IS |
| JNT-$Si_{32}$ | *P2_1/*c | 6.55 | 10.76 | 12.36 | 1.95 | 23.86 | 5.14 | 0.74 | IS |
| JOZ-$Si_{20}$ | *Pbcm* | 5.45 | 10.02 | 10.39 | 1.64 | 28.40 | 4.94 | 0.54 | IS |
| JRY-$Si_{24}$ | *I2_12_12_1* | 6.27 | 6.94 | 13.11 | 1.96 | 23.78 | 5.19 | 0.80 | IS |
| JSN-$Si_{16}$ | *P2/c* | 6.68 | 5.35 | 11.66 | 1.79 | 25.99 | 5.12 | 0.43 | IS |
| JSR-$Si_{96}$ | *Pa*-3 | 15.87 | 15.87 | 15.87 | 1.12 | 41.63 | 4.72 | 0.56 | IS |
| JST-$Si_{48}$ | *Pa*-3 | 12.24 | 12.24 | 12.24 | 1.22 | 38.17 | 4.66 | 0.46 | IS |
| JSW-$Si_{48}$ | *Pbca* | 7.49 | 12.44 | 12.86 | 1.87 | 24.94 | 5.11 | 0.96 | IS |
| KFI-$Si_{96}$ | *Im*-3*m* | 14.26 | 14.26 | 14.26 | 1.54 | 30.20 | 5.05 | 0.63 | IS |
| LAU-$Si_{24}$ | *C2/m* | 10.48 | 10.42 | 5.73 | 1.88 | 24.86 | 5.14 | 0.76 | IS |
| LEV-$Si_{54}$ | *R*-3*m* | 9.95 | 9.95 | 17.57 | 1.67 | 27.91 | 5.10 | 0.74 | DS |
| LIO-$Si_{36}$ | *P*-6*m*2 | 9.47 | 9.47 | 11.71 | 1.85 | 25.26 | 5.20 | 0.00 | SM |
| LIT-$Si_{24}$ | *Pnma* | 6.73 | 6.79 | 11.63 | 2.11 | 22.13 | 5.28 | 0.48 | DS |
| LOS-$Si_{24}$ | *P6_3/mmc* | 9.48 | 9.48 | 7.81 | 1.84 | 25.31 | 5.20 | 0.00 | SM |
| LOV-$Si_{18}$ | *P4_2/mmc* | 5.52 | 5.52 | 15.98 | 1.72 | 27.09 | 5.02 | 0.46 | IS |
| LTA-$Si_{24}$ | *Pm*-3*m* | 9.11 | 9.11 | 9.11 | 1.48 | 31.53 | 5.06 | 0.43 | IS |
| LTF-$Si_{108}$ | *P6_3/mmc* | 24.04 | 24.04 | 5.82 | 1.73 | 26.98 | 5.08 | 0.44 | IS |
| LTJ-$Si_{16}$ | *P4_12_12* | 7.00 | 7.00 | 7.85 | 1.94 | 24.05 | 5.17 | 0.93 | IS |
| LTL-$Si_{36}$ | *P6/mmm* | 14.09 | 14.09 | 5.87 | 1.66 | 28.01 | 5.06 | 0.00 | M |
| LTN-$Si_{768}$ | *Fd*-3*m* | 27.36 | 27.36 | 27.36 | 1.75 | 26.68 | 5.15 | 0.40 | DS |

| Si-phases | *SG* | *a* | *b* | *c* | *ρ* | $V_0$ | $E_c$ | $E_g$ | *P* |
|---|---|---|---|---|---|---|---|---|---|
| MAR-$Si_{72}$ | *P*6_3/*mmc* | 9.49 | 9.49 | 23.36 | 1.84 | 25.30 | 5.20 | 0.01 | DS |
| MAZ-$Si_{36}$ | *P*6_3/*mmc* | 14.02 | 14.02 | 5.80 | 1.70 | 27.39 | 5.08 | 0.58 | IS |
| MEI-$Si_{34}$ | *P*6_3/*m* | 10.12 | 10.12 | 12.15 | 1.47 | 31.68 | 5.08 | 0.95 | IS |
| MEL-$Si_{96}$ | *I*-4*m*2 | 15.35 | 15.35 | 10.35 | 1.84 | 25.41 | 5.25 | 0.94 | DS |
| MEP-$Si_{46}$ | *Pm*-3*n* | 10.23 | 10.23 | 10.23 | 2.01 | 23.26 | 5.36 | 1.31 | IS |
| MER-$Si_{32}$ | *I*4/*mmm* | 11.02 | 11.02 | 7.22 | 1.70 | 27.41 | 5.06 | 0.17 | DS |
| MFI-$Si_{96}$ | *Pnma* | 10.33 | 15.24 | 15.30 | 1.86 | 25.10 | 5.27 | 1.27 | IS |
| MFS-$Si_{36}$ | *Imm*2 | 5.74 | 11.08 | 14.12 | 1.87 | 24.96 | 5.21 | 0.74 | IS |
| MON-$Si_{16}$ | *I*4_1/*amd* | 5.46 | 5.46 | 13.00 | 1.92 | 24.24 | 5.23 | 1.11 | IS |
| MOR-$Si_{48}$ | *Cmcm* | 14.02 | 15.19 | 5.76 | 1.82 | 25.58 | 5.22 | 0.61 | IS |
| MOZ-$Si_{108}$ | *P*6/*mmm* | 24.12 | 24.12 | 5.85 | 1.71 | 27.31 | 5.07 | 0.00 | M |
| MSE-$Si_{112}$ | *P*4_2/*mnm* | 14.09 | 14.09 | 15.50 | 1.70 | 27.47 | 5.20 | 1.09 | DS |
| MSO-$Si_{90}$ | *R*-3*m* | 13.32 | 13.32 | 14.38 | 1.90 | 24.54 | 5.16 | 0.08 | IS |
| MTF-$Si_{44}$ | *C*2/*m* | 7.23 | 23.87 | 5.67 | 2.10 | 22.20 | 5.27 | 0.94 | IS |
| MTN-$Si_{136}$ | *Fd*-3*m* | 14.74 | 14.74 | 14.74 | 1.98 | 23.55 | 5.37 | 1.38 | DS |
| MTT-$Si_{24}$ | *Pmmn* | 3.86 | 8.58 | 16.82 | 2.01 | 23.21 | 5.31 | 0.47 | IS |
| MTW-$Si_{28}$ | *C*2/*m* | 19.12 | 3.97 | 9.34 | 1.93 | 24.11 | 5.29 | 0.56 | IS |
| MVY-$Si_{12}$ | *Pnnm* | 3.78 | 6.75 | 10.91 | 2.01 | 23.22 | 5.10 | 0.00 | M |
| MWF-$Si_{1440}$ | *Im*-3*m* | 34.33 | 34.33 | 34.33 | 1.66 | 28.16 | -5.06 | — | — |
| MWW-$Si_{72}$ | *P*6/*mmm* | 10.87 | 10.87 | 19.38 | 1.69 | 27.55 | 5.19 | 1.19 | IS |
| NAB-$Si_{10}$ | *I*-4*m*2 | 5.59 | 5.59 | 9.27 | 1.61 | 28.98 | 4.90 | 0.61 | IS |
| NAT-$Si_{20}$ | *I*4_1/*amd* | 10.45 | 10.45 | 4.99 | 1.71 | 27.23 | 5.12 | 1.35 | IS |
| NES-$Si_{34}$ | *Fmmm* | 10.08 | 10.08 | 11.06 | 1.79 | 26.00 | 5.25 | 0.74 | IS |
| NON-$Si_{22}$ | *Fmmm* | 7.81 | 7.81 | 9.97 | 2.00 | 23.35 | 5.33 | 1.13 | IS |
| NPO-$Si_{6}$ | *P*6_3/*mmc* | 6.94 | 6.94 | 3.91 | 1.72 | 27.19 | 5.10 | 0.16 | DS |
| NPT-$Si_{36}$ | *Pm*-3*m* | 10.92 | 10.92 | 10.92 | 1.29 | 36.17 | 4.79 | 0.38 | DS |
| NSI-$Si_{12}$ | *C*2/*m* | 10.74 | 3.87 | 6.56 | 2.13 | 21.89 | 5.34 | 0.63 | IS |
| OBW-$Si_{76}$ | *I*4/*mmm* | 11.01 | 11.01 | 24.87 | 1.18 | 39.66 | 4.73 | 0.34 | IS |
| OFF-$Si_{18}$ | *P*-6*m*2 | 9.93 | 9.93 | 5.87 | 1.68 | 27.84 | 5.09 | 0.03 | IS |
| OKO-$Si_{68}$ | *C*2/*m* | 18.59 | 10.72 | 9.45 | 1.78 | 26.22 | 5.22 | 0.97 | IS |
| OSI-$Si_{32}$ | *I*4/*mmm* | 13.97 | 13.97 | 3.98 | 1.93 | 24.17 | 5.25 | 0.14 | IS |
| OSO-$Si_{9}$ | *P*6_222 | 8.34 | 8.34 | 5.96 | 1.17 | 39.83 | 4.67 | 0.09 | IS |
| OWE-$Si_{16}$ | *Pmma* | 5.47 | 6.79 | 11.19 | 1.80 | 25.95 | 5.11 | 0.44 | IS |

| **Si-phases** | ***SG*** | ***a*** | ***b*** | ***c*** | ***ρ*** | $V_0$ | $E_c$ | $E_g$ | ***P*** |
|---|---|---|---|---|---|---|---|---|---|
| PAR-$Si_{16}$ | *P*-1 | 6.29 | 7.91 | 9.01 | 1.83 | 25.52 | 5.01 | 0.47 | IS |
| PAU-$Si_{672}$ | *Im-3m* | 26.73 | 26.73 | 26.73 | 1.64 | 28.474 | -5.05 | — | — |
| PCR-$Si_{60}$ | *C2/m* | 15.21 | 10.75 | 9.41 | 1.99 | 23.48 | 5.29 | 0.99 | IS |
| PHI-$Si_{32}$ | *Cmcm* | 7.34 | 10.91 | 10.98 | 1.70 | 27.43 | 5.07 | 0.50 | IS |
| PON-$Si_{24}$ | *Pca*2_1 | 6.86 | 7.09 | 12.47 | 1.84 | 25.29 | 5.15 | 0.75 | IS |
| POS-$Si_{64}$ | *P4_2/mnm* | 14.58 | 14.58 | 8.81 | 1.59 | 29.24 | 5.14 | 0.83 | IS |
| PSI-$Si_{144}$ | *Cmce* | 6.38 | 17.22 | 29.34 | 2.08 | 22.38 | 5.30 | 0.40 | IS |
| PUN-$Si_{36}$ | *Pbcn* | 6.72 | 11.30 | 15.05 | 1.47 | 31.76 | 4.91 | 0.48 | IS |
| RHO-$Si_{48}$ | *Im-3m* | 11.52 | 11.52 | 11.52 | 1.46 | 31.87 | 5.02 | 0.58 | IS |
| RON-$Si_{56}$ | *C2/c* | 14.93 | 13.17 | 7.06 | 2.14 | 21.79 | 4.96 | 0.00 | M |
| RRO-$Si_{18}$ | *P2/c* | 5.67 | 6.66 | 13.14 | 1.84 | 25.28 | 5.18 | 0.69 | IS |
| RSN-$Si_{36}$ | *Cmmm* | 5.60 | 31.59 | 5.49 | 1.73 | 27.01 | 4.98 | 0.56 | IS |
| RTE-$Si_{24}$ | *C2/m* | 10.68 | 10.55 | 5.66 | 1.79 | 26.07 | 5.19 | 1.06 | IS |
| RTH-$Si_{32}$ | *C2/m* | 7.55 | 16.07 | 7.46 | 1.66 | 28.18 | 5.15 | 0.76 | IS |
| RUT-$Si_{36}$ | *C2/m* | 10.22 | 10.24 | 9.43 | 1.87 | 24.97 | 5.24 | 0.93 | IS |
| RWR-$Si_{32}$ | *I4_1/amd* | 5.94 | 5.94 | 20.85 | 2.03 | 23.00 | 5.22 | 0.50 | DS |
| RWY-$Si_{24}$ | *C2/m* | 12.49 | 12.49 | 12.49 | 0.75 | 62.55 | 4.38 | 0.00 | M |
| SAF-$Si_{64}$ | *Ibam* | 6.38 | 11.23 | 21.35 | 1.95 | 23.91 | 5.27 | 0.55 | IS |
| SAO-$Si_{56}$ | *I*-4*m*2 | 10.66 | 10.66 | 16.55 | 1.39 | 33.60 | 5.03 | 0.53 | IS |
| SAS-$Si_{32}$ | *I4/mmm* | 10.92 | 10.92 | 7.76 | 1.61 | 28.90 | 5.10 | 0.27 | IS |
| SAT-$Si_{72}$ | *R*-3*m* | 9.76 | 9.76 | 23.54 | 1.73 | 27.00 | 5.12 | 0.10 | DS |
| SAV-$Si_{48}$ | *P4/nmm* | 14.21 | 14.21 | 7.14 | 1.55 | 30.02 | 5.06 | 0.59 | IS |
| SBE-$Si_{128}$ | *I4/mmm* | 14.31 | 14.31 | 21.45 | 1.37 | 34.14 | 5.03 | 0.73 | IS |
| SBN-$Si_{10}$ | *P6_3/mmc* | 5.56 | 5.56 | 10.58 | 1.65 | 28.31 | 4.99 | 0.67 | IS |
| SBS-$Si_{96}$ | *P6_3/mmc* | 13.29 | 13.29 | 21.60 | 1.36 | 34.33 | 5.03 | 0.81 | IS |
| SBT-$Si_{144}$ | *R*-3*m* | 13.31 | 13.31 | 32.38 | 1.35 | 34.48 | 5.04 | 0.76 | IS |
| SEW-$Si_{66}$ | *Pmmn* | 8.85 | 11.05 | 18.74 | 1.68 | 27.77 | 5.16 | 0.72 | IS |
| SFE-$Si_{14}$ | *P2_1/m* | 8.62 | 3.84 | 10.92 | 1.83 | 25.54 | 5.23 | 0.08 | IS |
| SFF-$Si_{32}$ | *P2_1/m* | 5.70 | 17.08 | 8.89 | 1.73 | 26.92 | 5.19 | 1.15 | IS |
| SFG-$Si_{74}$ | *Pmma* | 9.63 | 9.95 | 19.32 | 1.86 | 25.02 | 5.26 | 0.66 | IS |
| SFH-$Si_{64}$ | *Cmcm* | 4.26 | 25.65 | 15.53 | 1.76 | 26.56 | 5.12 | 0.00 | M |
| SFN-$Si_{32}$ | *C2/m* | 19.21 | 3.95 | 11.54 | 1.75 | 26.72 | 5.23 | 0.15 | IS |
| SFO-$Si_{32}$ | *C2/m* | 17.25 | 10.54 | 5.38 | 1.54 | 30.32 | 5.09 | 0.93 | IS |

| Si-phases | *SG* | *a* | *b* | *c* | *ρ* | $V_0$ | $E_c$ | $E_g$ | *P* |
|---|---|---|---|---|---|---|---|---|---|
| SFS-$Si_{56}$ | *P*2_1/*m* | 9.47 | 15.38 | 10.83 | 1.72 | 27.13 | 5.20 | 0.87 | IS |
| SFW-$Si_{108}$ | *R*-3*m* | 10.53 | 10.53 | 33.52 | 1.56 | 29.82 | 5.07 | 0.75 | IS |
| SGT-$Si_{64}$ | *I*4_1/*amd* | 7.81 | 7.81 | 26.70 | 1.83 | 25.46 | 5.26 | 1.03 | DS |
| SIV-$Si_{64}$ | *Cmcm* | 7.40 | 10.86 | 21.78 | 1.70 | 27.37 | 5.07 | 0.63 | IS |
| SOD-$Si_{12}$ | *Im*-3*m* | 6.73 | 6.73 | 6.73 | 1.84 | 25.40 | 5.20 | 0.14 | IS |
| SOF-$Si_{40}$ | *C*2/*c* | 15.44 | 9.75 | 7.61 | 1.71 | 27.32 | 5.07 | 0.82 | IS |
| SOR-$Si_{48}$ | *Cmmm* | 13.60 | 15.81 | 5.87 | 1.77 | 26.33 | 5.11 | 0.06 | IS |
| SOS-$Si_{24}$ | *Pmna* | 5.81 | 7.47 | 15.91 | 1.62 | 28.77 | 4.86 | 0.65 | IS |
| SSF-$Si_{54}$ | *P*6/*mmm* | 13.17 | 13.17 | 9.59 | 1.75 | 26.67 | 5.19 | 0.35 | IS |
| SSY-$Si_{28}$ | *Pmmn* | 3.85 | 10.92 | 17.02 | 1.83 | 25.55 | 5.23 | 0.06 | IS |
| STF-$Si_{32}$ | *C*2/*m* | 10.81 | 14.18 | 5.70 | 1.73 | 26.92 | 5.20 | 1.06 | IS |
| STI-$Si_{18}$ | *Fmmm* | 8.50 | 8.50 | 8.62 | 1.75 | 26.60 | 5.14 | 0.50 | IS |
| STT-$Si_{64}$ | *P*2_1/*c* | 10.11 | 16.77 | 12.87 | 1.71 | 27.33 | 5.17 | 1.02 | IS |
| STW-$Si_{60}$ | *P*6_122 | 9.40 | 9.40 | 22.40 | 1.63 | 28.60 | 5.05 | 0.72 | IS |
| SVR-$Si_{92}$ | *Cc* | 15.77 | 10.25 | 14.92 | 1.88 | 24.75 | 5.16 | 0.44 | IS |
| SVV-$Si_{56}$ | *C*2/*m* | 10.24 | 10.24 | 13.77 | 1.82 | 25.67 | 5.20 | 0.71 | DS |
| SWY-$Si_{72}$ | *P*6_3/*mmc* | 9.96 | 9.96 | 23.34 | 1.68 | 27.79 | 5.09 | 0.00 | M |
| SZR-$Si_{36}$ | *Cmmm* | 10.99 | 13.95 | 5.80 | 1.89 | 24.65 | 5.19 | 0.32 | IS |
| TER-$Si_{80}$ | *Cmcm* | 7.50 | 18.05 | 15.22 | 1.81 | 25.74 | 5.21 | 0.81 | IS |
| THO-$Si_{10}$ | *Pmma* | 4.96 | 5.25 | 10.47 | 1.71 | 27.22 | 5.12 | 1.03 | DS |
| TOL-$Si_{72}$ | *P*-3*m*1 | 9.47 | 9.47 | 23.41 | 1.85 | 25.24 | 5.20 | 0.00 | SM |
| TON-$Si_{24}$ | *Cmcm* | 10.87 | 13.29 | 3.86 | 2.01 | 23.20 | 5.31 | 0.39 | IS |
| TSC-$Si_{384}$ | *Fm*-3*m* | 23.80 | 23.80 | 23.80 | 1.33 | 35.11 | 5.02 | 0.76 | IS |
| TUN-$Si_{192}$ | *C*2/*m* | 21.80 | 15.32 | 14.93 | 1.80 | 25.97 | 5.22 | 1.20 | IS |
| UEI-$Si_{12}$ | *Fmm*2 | 6.86 | 6.86 | 8.04 | 1.88 | 24.76 | 5.13 | 0.58 | IS |
| UFI-$Si_{64}$ | *I*4/*mmm* | 9.40 | 9.40 | 21.38 | 1.58 | 29.53 | 5.13 | 0.87 | IS |
| UOS-$Si_{24}$ | *Pmma* | 5.88 | 7.01 | 14.68 | 1.85 | 25.23 | 5.13 | 0.04 | IS |
| UOV-$Si_{176}$ | *Amm*2 | 16.96 | 30.15 | 9.68 | 1.66 | 28.12 | 5.16 | 0.82 | IS |
| UOZ-$Si_{40}$ | *P*4/*nnc* | 6.87 | 6.87 | 21.77 | 1.82 | 25.68 | 5.01 | 0.00 | M |
| USI-$Si_{40}$ | *C*2/*m* | 16.79 | 10.17 | 7.37 | 1.55 | 30.06 | 5.05 | 0.18 | DS |
| UTL-$Si_{76}$ | *C*2/*m* | 22.93 | 10.75 | 9.46 | 1.57 | 29.72 | 5.17 | 0.98 | IS |
| UWY-$Si_{60}$ | *Pmmm* | 8.78 | 9.72 | 19.28 | 1.70 | 27.40 | 5.18 | 0.77 | IS |
| VET-$Si_{17}$ | *P*-4 | 10.03 | 10.03 | 3.88 | 2.03 | 22.96 | 5.30 | 0.87 | IS |

| Si-phases | *SG* | *a* | *b* | *c* | *ρ* | $V_0$ | $E_c$ | $E_g$ | *P* |
|---|---|---|---|---|---|---|---|---|---|
| VFI-$Si_{36}$ | *P6_3/mcm* | 14.47 | 14.47 | 6.33 | 1.46 | 31.88 | 5.15 | 0.00 | SM |
| VNI-$Si_{60}$ | *P4_2/ncm* | 7.67 | 7.67 | 26.16 | 1.82 | 25.66 | 5.08 | 1.06 | IS |
| VSV-$Si_{36}$ | *I4_1/amd* | 5.52 | 5.52 | 31.73 | 1.74 | 26.87 | 5.05 | 0.76 | IS |
| WEI-$Si_{20}$ | *Cccm* | 7.78 | 9.31 | 8.21 | 1.57 | 29.72 | 4.86 | 0.33 | IS |
| WEN-$Si_{20}$ | *P*-62*m* | 10.30 | 10.30 | 5.91 | 1.72 | 27.15 | 5.00 | 0.00 | M |
| YFI-$Si_{120}$ | *Cmmm* | 13.88 | 24.48 | 9.58 | 1.71 | 27.22 | 5.19 | 0.98 | IS |
| YUG-$Si_{16}$ | *C*2/*m* | 7.49 | 10.70 | 5.11 | 1.93 | 24.14 | 5.21 | 1.00 | IS |
| ZON-$Si_{32}$ | *Pbcm* | 5.36 | 11.71 | 13.29 | 1.79 | 25.98 | 5.12 | 0.38 | IS |

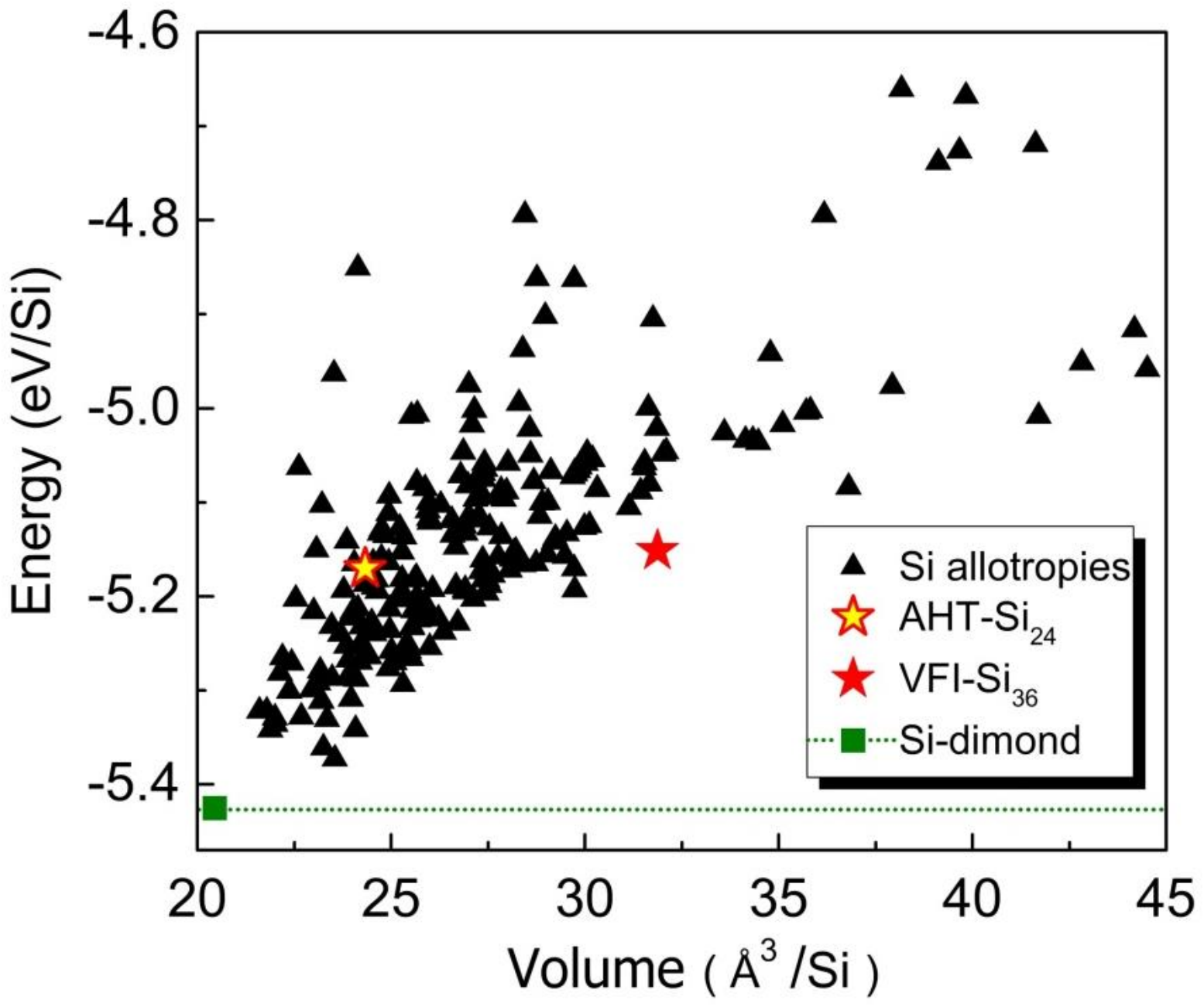

FIG. S2. Energy per Si atom as a function of volume for all the optimized Si-allotropes in Table SI. The topological nodal-line semimetallic AHT-$Si_{24}$ and VFI-$Si_{36}$ are marked by five-pointed stars in different color. The energy of *d*-Si is also presented as green dashed line for comparison.

## III. Structure and Stability of AHT-$Na_4Si_{24}$

On the basis of the structural character of AHT-$Si_{24}$, we constructed a hypothetical Na-Si system as the possible precursor, termed as AHT-$Na_4Si_{24}$. Upon fully optimization, AHT-$Na_4Si_{24}$ can well maintain the crystal structure of pristine AHT-$Si_{24}$ at ambient pressure [see Fig. S3(a)]. The key parameters are listed in Table SII. For comparison, the results of previously reported precursors, *e.g.*, *P*/6*m*-$NaSi_6$ and *Cmcm*-$Na_4Si_{24}$, are also presented. Notably, the cohesive energy of AHT-$Na_4Si_{24}$ is larger than that of *P*/6*m*-$NaSi_6$ by 0.9 eV per $NaSi_6$, suggesting that when AHT-$Na_4Si_{24}$ is fabricated, it would be even more stable than *P*/6*m*-$NaSi_6$ at ambient pressure. Furthermore, the stability of AHT-$Na_4Si_{24}$ at ambient condition has been confirmed by the first-principle molecular dynamics (FPMD) simulations at 300 K [see Fig. S3(b)].

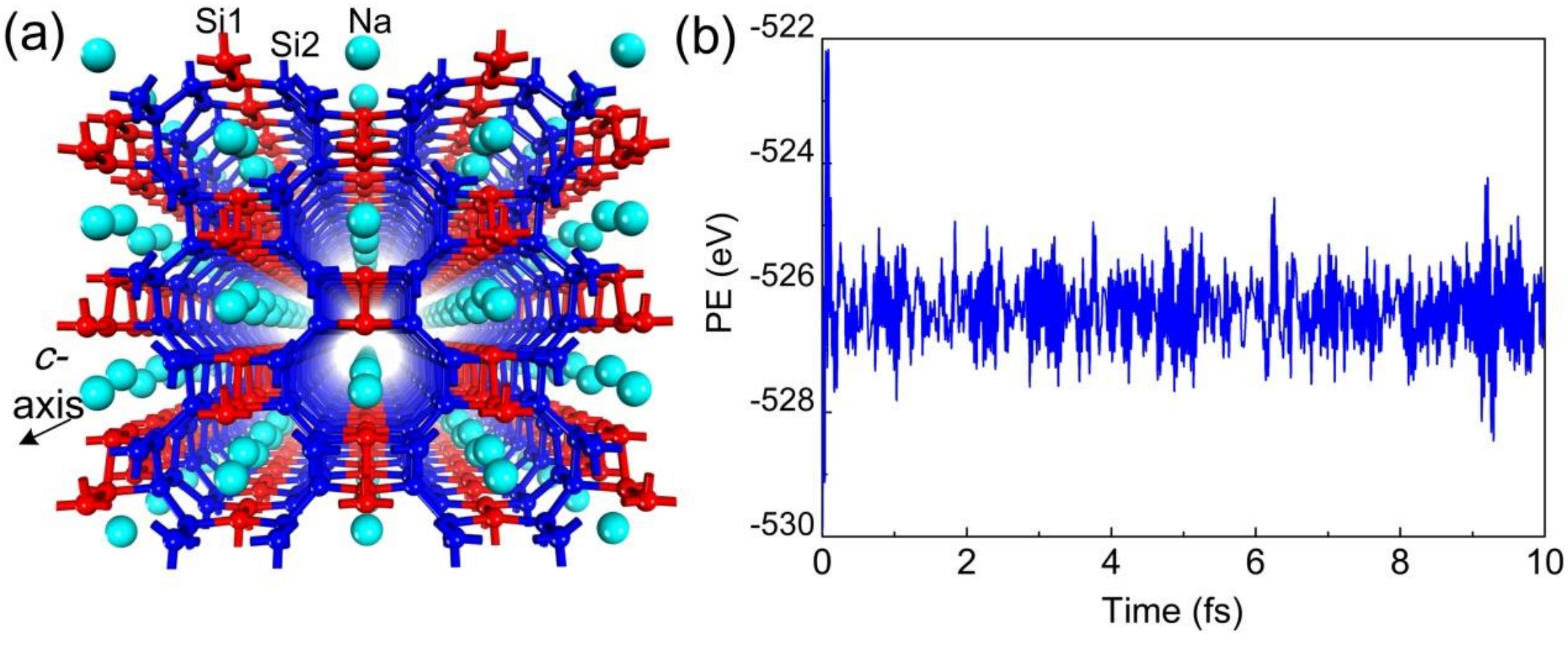

FIG. S3 (a) A perspective view from crystallographic *c*-axis for the optimized AHT-$Na_4Si_{24}$ in *Cmcm* symmetry. (b) Total potential energy (PE) fluctuation for AHT-$Na_4Si_{24}$ during the FPMD simulations at 300 K and zero pressure. After 10 ps with a MD time step of 1 fs, no structural destruction is found, except for some thermal-induced fluctuations, suggesting that the AHT-$Na_4Si_{24}$ is stable at ambient conditions.

Table SII. Calculated equilibrium lattice parameters (*a*, *b* and *c* in Å), density *ρ* (g/cm$^3$), volume (Å$^3$/$NaSi_6$), bulk modulus $B_0$ (GPa), and cohesive energy $E_c$ (eV/$NaSi_6$) for AHT-$Na_4Si_{24}$, and the reported *P*6/*m*-$NaSi_6$ and *Cmcm*-$Na_4Si_{24}$ phases. The experimental data are highlighted with bold type.

| Phases | *a* | *b* | *c* | *ρ* | $V_0$ | $B_0$ | $E_c$ |
|---|---|---|---|---|---|---|---|
| AHT-$Na_4Si_{24}$ | 12.10 | 7.71 | 6.36 | 2.145 | 148.24 | 62 | 32.86 |
| P6/*m*-$NaSi_6$ | 7.09 | | 2.54 | 2.872 | 110.71 | 72 | 31.96 |
| | 7.04[1] | | 2.57[1] | | | | |
| *Cmcm*-$Na_4Si_{24}$ | 4.03 | 10.71 | 12.47 | 2.358 | 134.83 | 74 | 33.63 |
| | 4.12[2] | 10.58[2] | 12.28[2] | | | | |
| | **4.10[2]** | **10.57[2]** | **12.26[2]** | | | | |

## IV Energy *vs.* Volume for Some Compact Si Allotropes.

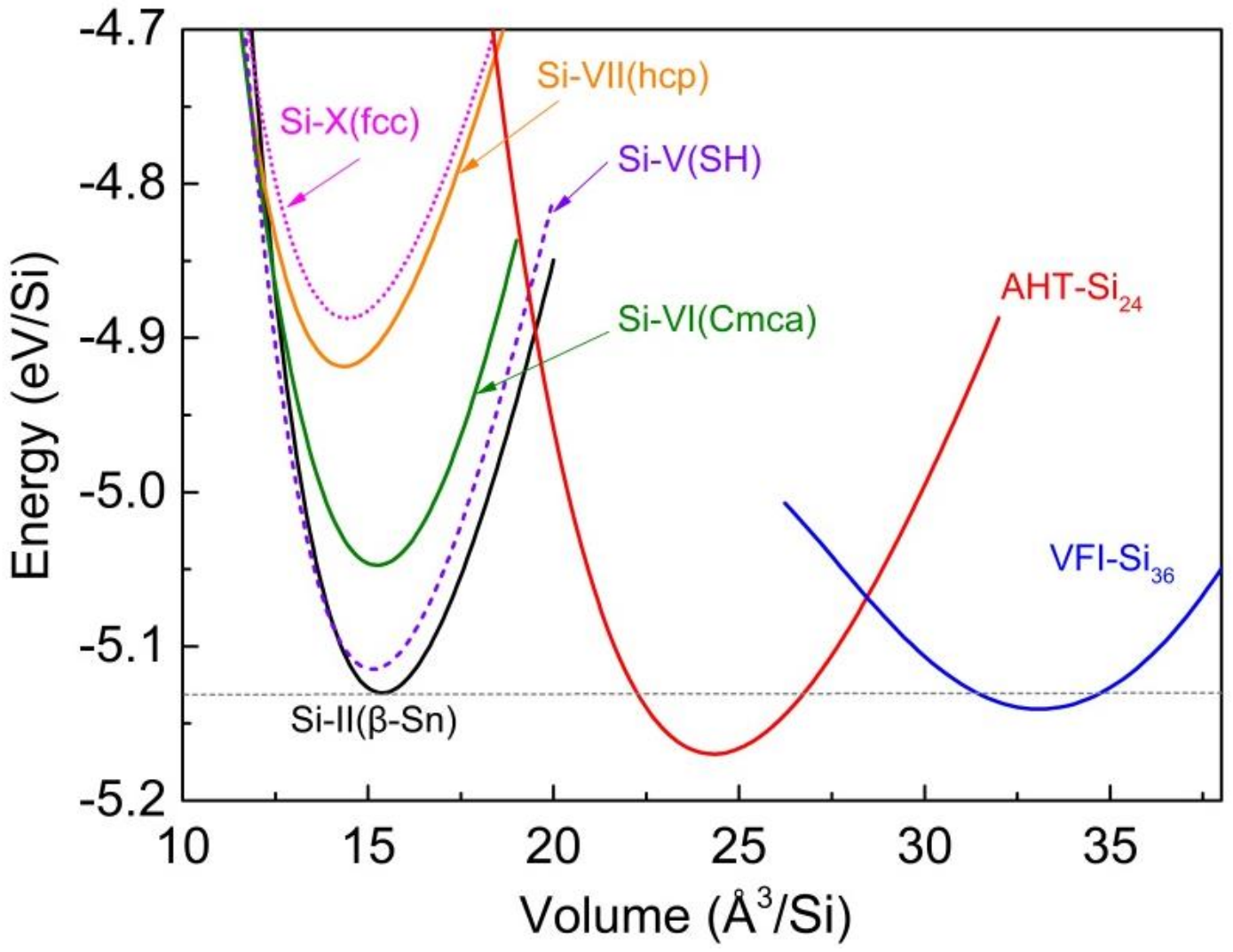

FIG. S4 Energy per Si atom as a function of volume for AHT-$Si_{24}$ and VFI-$Si_{36}$ compared to some synthesized compact Si allotropes at zero pressure.

## V. Electronic Band Structures from HSE06 calculation

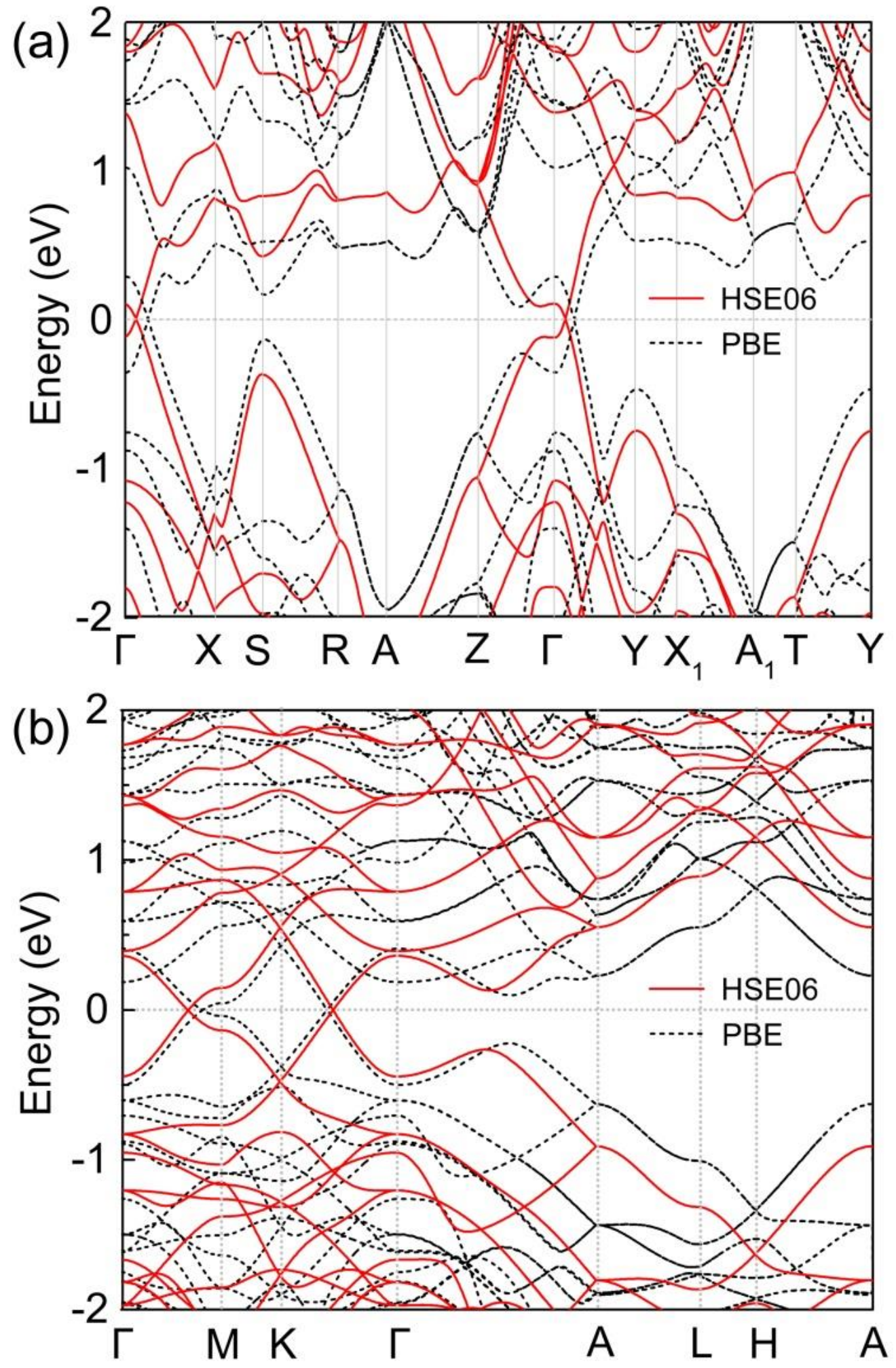

FIG. S5. The electronic band structures (red solid line) calculated using HSE06 method for (a) AHT-$Si_{24}$ and (b) VFI-$Si_{36}$. For comparison, the corresponding bands from PBE calculation are also displayed (black dashed line). Although the positions of most bands show relative shifts with respect to those from PBE method, the linear Dirac crossings between HVB and LCB are well preserved in the HSE06 calculations.

## VI. Atomic Structure and Stability of VFI-$Si_{36}$

The optimized structure of VFI-$Si_{36}$ is shown in Fig. S6 (a). Like in AHT-$Si_{24}$, there are two kinds of crystallographically inequivalent atoms, denoted by Si1 and Si2, and all the atoms are four-fold coordinated with every Si1 atom connected to two Si1 and two Si2 atoms, and each Si2 bonded with one Si1 and three Si2 atoms. The average bond angle is calculated to be 107.44°, comparable to 109.47° in *d*-Si phase. Furthermore, along *c*-axis two types of open channels can be found [see Fig. 6S (b)]; the larger open channels are constructed by 18-membered Si rings, and the smaller ones are built from 6-membered Si rings.

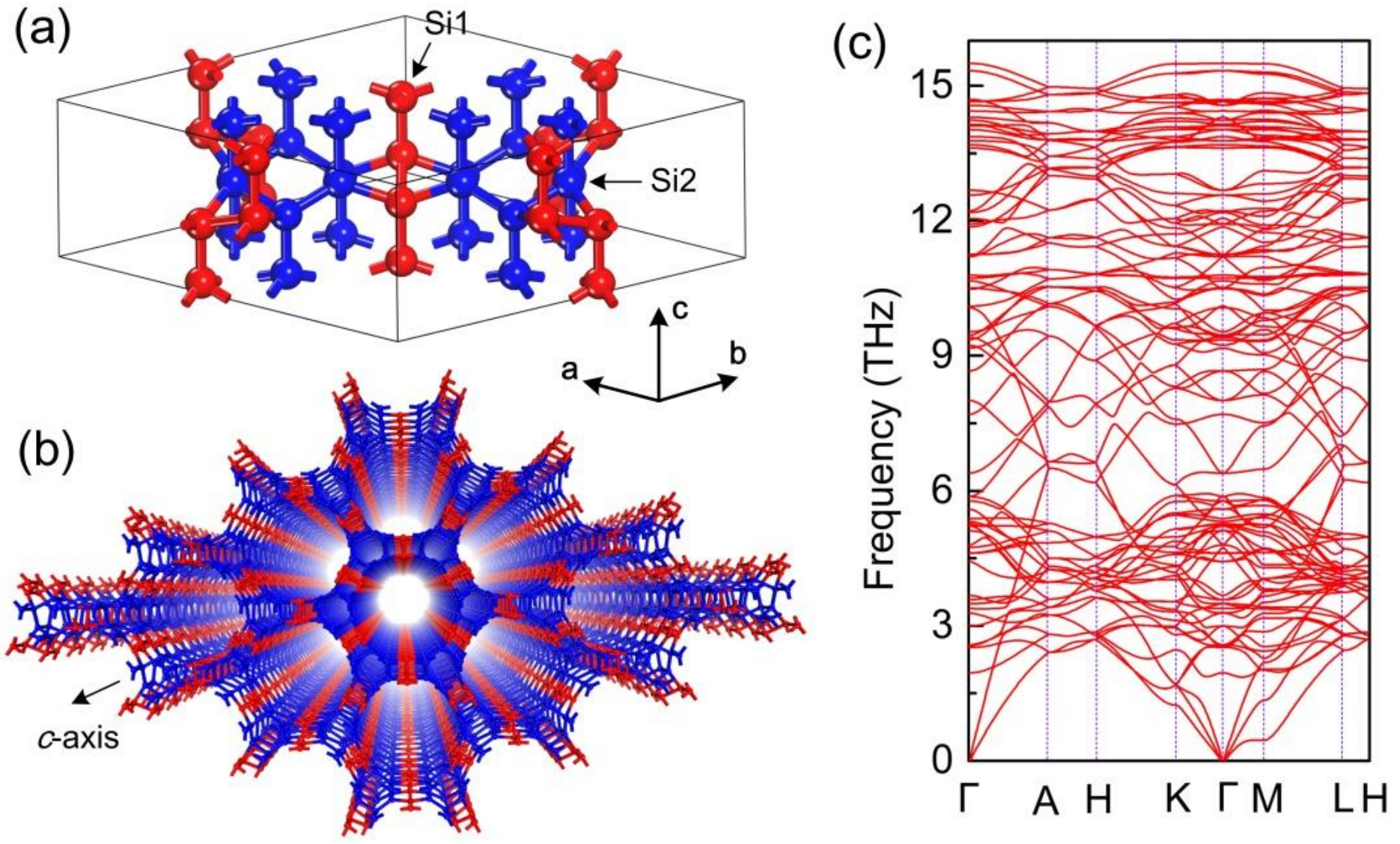

FIG. S6. (a) The primitive cell of VFI-$Si_{36}$ in $P6_3/mcm$ ($D_{6h}^3$) space group with lattice constants $a = b = 14.47$ Å and $c = 6.33$ Å, consisting two types of crystallographically inequivalent atoms: Si1 (red) and Si2 (blue) at 12*k* (0.4219, 0.4219, 0.5645) and 24*l* (0.1783, 0.5106, 0.9360) Wyckoff positions, respectively. (b) A perspective view from crystallographic *c*-axis for the atomic structure of VFI-$Si_{36}$. (c) Phonon dispersion of VFI-$Si_{36}$, suggesting that the structure is dynamical stable.

The cohesive energy of VFI-$Si_{36}$ is 5.15 eV/atom, which is slightly less than that of AHT-$Si_{24}$ by 0.02 eV/atom. However, it is still more stable than the prominent open-framework phases *P*/6*m*-$Si_6$ (5.07 eV/atom) [1] and $Si_{20}$-T [3] (5.14 eV/eV), and some synthesized compact phases [see Fig. S4] at ambient pressure. The dynamic stability of VFI-$Si_{36}$ is confirmed by the computed phonon spectrum in Fig. S6 (c), showing the absence of imaginary frequency over the entire BZ. Furthermore, its independent elastic constants are calculated to be $C_{11}$ = 77 GPa, $C_{12}$ = 35 GPa, $C_{13}$ = 21 GPa, $C_{33}$ = 124 GPa and $C_{44}$ = 21 GPa, which meet the criteria for mechanical stability given by $C_{44} > 0$, $| C_{11} | > C_{12}$, and $C_{11} + 2 C_{12} C_{33} > 2 C_{13}^2$ for a hexagonal crystal [4].

## VII. Electronic Band Structure of VFI-$Si_{36}$

As shown in Fig. S7(a), on the Fermi level the HVB and LCB of VFI-$Si_{36}$ linearly cross at two asymmetric Dirac nodal points, namely D1 in the Γ-M and D2 in Γ-K path. From the orbital and atom decomposed band structures near D1 and D2 [see Fig. S7(c) and 2(d)], one can find that the crossing bands are mainly originated from the $p_{xy}$ states of both Si1 and Si2 atoms, termed as $\left|\mathrm{Si1}+\mathrm{Si2},\, p_{xy}\right\rangle$, and the $s$ and $p_{xy}$ hybridized states of Si2 atoms denoted as $\left|\mathrm{Si2},\, s{+}p_{xy}\right\rangle$. Hence, the band inversion between $\left|\mathrm{Si1}+\mathrm{Si2},\, p_{xy}\right\rangle$ and $\left|\mathrm{Si2},\, s{+}p_{xy}\right\rangle$ states should be responsible for the band crossings.

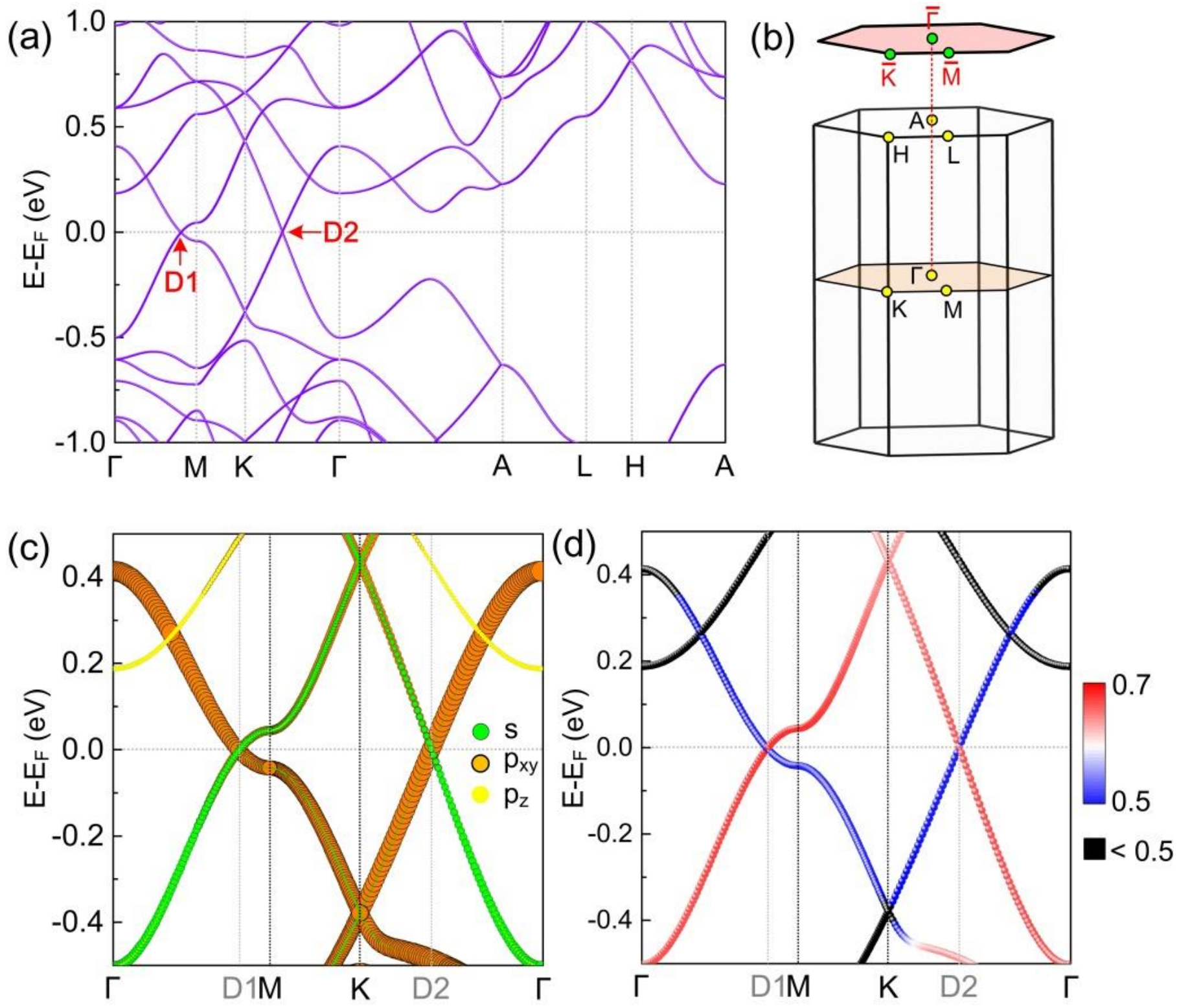

FIG. S7. (a) Bulk electronic band structure of VFI-$Si_{36}$. (b) The corresponding bulk and (001)-surface BZ together with the high-symmetry ***k***-points. (c) Orbital decomposed linear crossing bands around the nodal points. (d) Colorful band structure near the nodal points weighted by the contribution ratio, *i.e.*, $\eta = W_{Si2} / (W_{Si1} + W_{Si2})$ , where W denotes the weight of the corresponding atom.

### VIII. Nodal Line and Flat Surface Bands of VFI-$Si_{36}$

The calculated Fermi surface and 3D band structure of VFI-$Si_{36}$ show that the HVB and LCB cross with each other along a closed loop inside the BZ [see Fig. S8(a)]. Thus, similar to AHT-$Si_{24}$, VFI-$Si_{36}$ is also a topological node-line semimetal. Owing to the hexagonal crystalline symmetry, the nodal Dirac points form a "rounded hexagon" nodal line within the $k_z = 0$ plane [see Fig. S8 (a)].

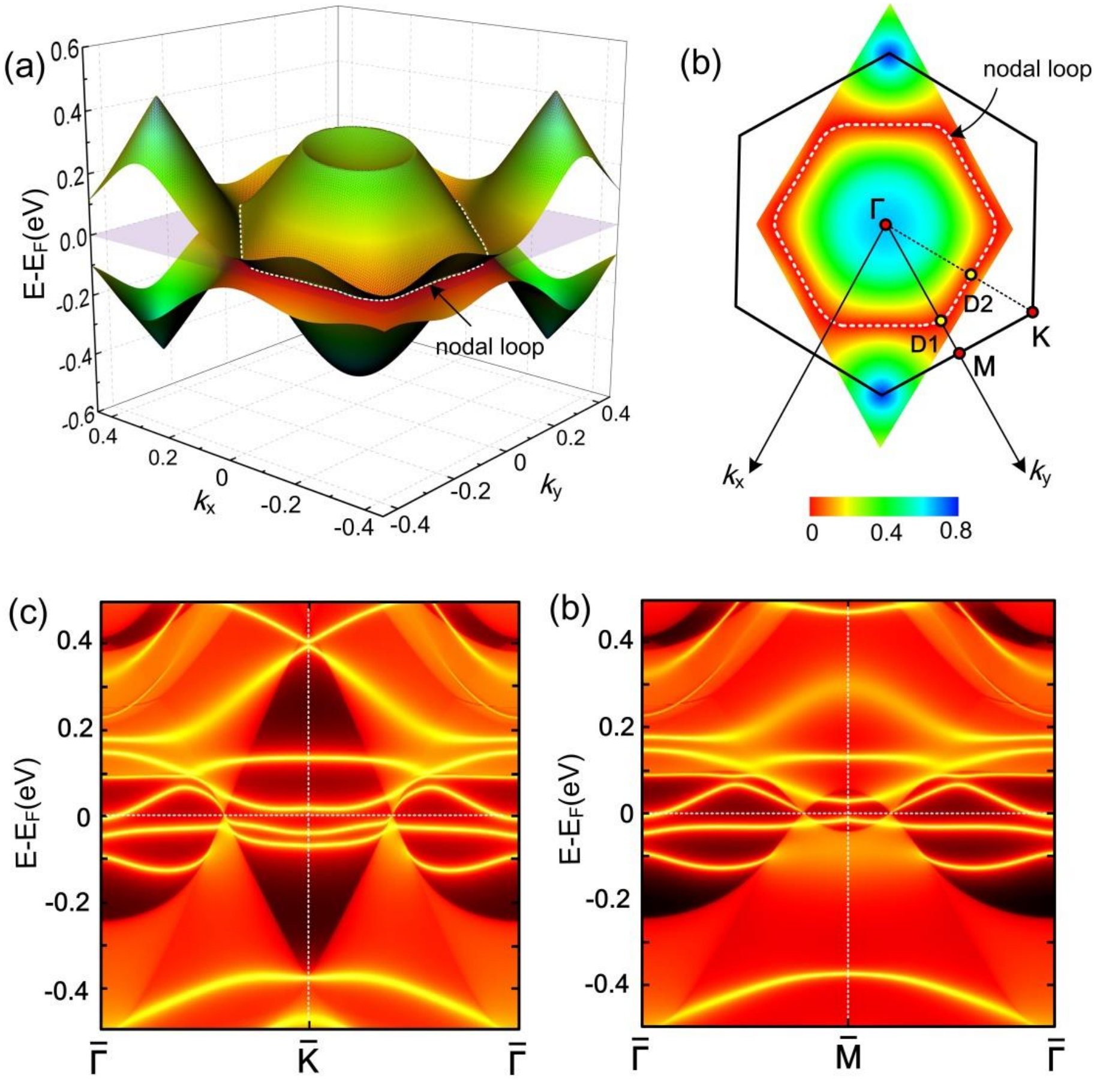

FIG. S8. (a) 3D band structure of the HVB and LCB for VFI-$Si_{24}$. (b) The colorized contour plot of the energy band gap between HVB and LCB in the $k_z$=0 plane. The (001)-surface states along (c) $\bar{\Gamma}-\bar{K}$ and (d) $\bar{\Gamma}-\bar{M}$ directions within the $\bar{\Gamma}-\bar{K}-\bar{M}-\bar{\Gamma}$ plane [see Fig. S7 (b)].

Since the closed nodal line is protected by space inversion, time-reversal, and SU(2) spin-rotation symmetries, VFI-$Si_{36}$ can be classified as a type-B topological node-line semimetal [5]. The topological invariant $\zeta_1$ is calculated to be 1, implying that the band crossings along the nodal loop are non-accidental. By considering SOC, an exceptionally small band gap of 0.6 meV along Γ-M and 1.2 meV along Γ-K directions are estimated. In other words, when the temperature is larger than 14 K, the weak SOC would not alter the semimetallic properties of VFI-$Si_{36}$.

To further explore the topological properties of VFI-$Si_{36}$, we calculated The (001)-surface states of VFI-$Si_{36}$ along $\overline{\Gamma}-\overline{\mathrm{K}}$ and $\overline{\Gamma}-\overline{\mathrm{M}}$ directions, as shown in Figs. S8(c) and S8(d), respectively. The results show that around the Fermi energy, there two kinds of flat surface bands, which nestle either inside or outside of the rounded-hexagon nodal loop. In both $\overline{\Gamma}-\overline{\mathrm{K}}$ and $\overline{\Gamma}-\overline{\mathrm{M}}$ directions, the coupling surface states opens a gap with two bands located below and above the Fermi level, respectively. In experiment, these surface states should be detectable by photoelectron spectroscopy. Moreover, it is anticipated that such flat surface bands could induce the desirable high-temperature surface superconductivity [6,7].

## IX. Simulated XRD Patterns

To guide the experimental observation of AHT-$Si_{24}$ and VFI-$Si_{36}$, their X-ray diffraction (XRD) patterns were simulated. The results are plotted in Fig. S9. For AHT-$Si_{24}$, one can see two strong peaks at $2\theta = 13.9°$ and 28°, corresponding to the AHT-$Si_{24}$ (110) and (002) diffraction, respectively. As for VFI-$Si_{36}$, the (001) and (002) diffraction peaks can be found at $2\theta = 7.1°$ and 14.2°, respectively.

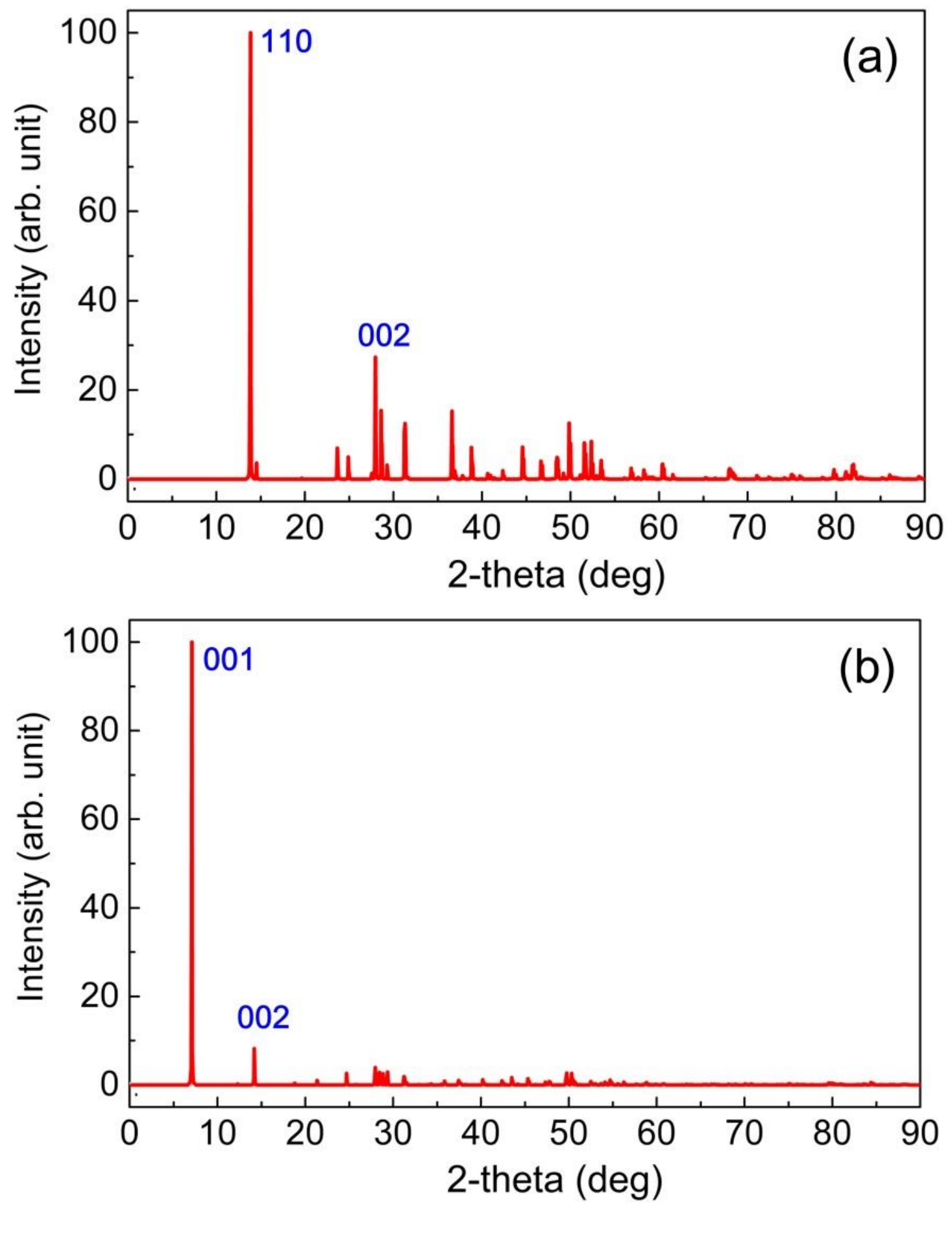

FIG. S9. Simulated XRD patterns for (a) AHT-$Si_{24}$ and (b) VFI-$Si_{36}$. The X-ray wavelength is 1.54056 Å with a Cu source.